\documentclass[fleqn,usenatbib]{mnras}

\usepackage{newtxtext,newtxmath}
\usepackage[T1]{fontenc}
\usepackage{ae,aecompl}

\usepackage{graphicx}	% Including figure files
\usepackage{amsmath}	% Advanced maths commands
\usepackage{amsbsy}	% Extra maths symbols

\newcommand*\Btheta{\ensuremath{\boldsymbol\theta}}

\title[The star-forming galaxy contribution to the EGB]{Characterising the signatures of star-forming galaxies in the extra-galactic $\gamma$-ray background}

\author[Owen, Lee \& Kong]{
Ellis R. Owen$^{1,2,3}$\thanks{E-mail: erowen@gapp.nthu.edu.tw (ERO)}, 
Khee-Gan Lee$^{4}$, Albert K. H. Kong$^{1}$%, 4}
\\
$^{1}$Institute of Astronomy, National Tsing Hua University, Hsinchu, Taiwan (ROC) \\
$^{2}$Center for Informatics and Computation in Astronomy, National Tsing Hua University, Hsinchu, Taiwan (ROC)\\
$^{3}$Mullard Space Science Laboratory, University College London, Holmbury St. Mary, Dorking, Surrey, RH5 6NT, United Kingdom\\
$^{4}$Kavli Institute for the Physics and Mathematics of the Universe, The University of Tokyo, Kashiwa, Chiba 277-8583, Japan
}

\date{Accepted XXX. Received YYY; in original form ZZZ}

\pubyear{2020}

\begin{document}
\label{firstpage}
\pagerange{\pageref{firstpage}--\pageref{lastpage}}
\maketitle

\begin{abstract}
Galaxies experiencing intense star-formation episodes 
are expected to be rich in energetic cosmic rays (CRs). 
These CRs undergo hadronic interactions with the interstellar gases of their host to drive $\gamma$-ray emission, 
which has already been detected from several nearby starbursts. 
Unresolved $\gamma$-ray emission from more distant star-forming galaxies (SFGs) is expected to contribute to 
the extra-galactic $\gamma$-ray background (EGB). However, despite the wealth of high-quality all-sky data 
from the \textit{Fermi}-LAT $\gamma$-ray space telescope collected over more than a decade of operation, 
the exact contribution of such SFGs to the EGB remains unsettled. 
We investigate the high-energy $\gamma$-ray emission from SFGs up to redshift $z=3$ above a GeV, and assess the contribution they can make to the EGB. We show 
the $\gamma$-ray emission spectrum
 from a SFG population can be determined from just a small number of key parameters, from which we model a range of possible EGB realisations. We demonstrate that populations of SFGs leave anisotropic signatures in the EGB, and that these can be accessed using the spatial power spectrum. 
Moreover, we show that such signatures will be accessible with ongoing operation of current $\gamma$-ray instruments, and detection prospects will be greatly improved by the next generation of $\gamma$-ray observatories, in particular the Cherenkov Telescope Array.
\end{abstract}
\begin{keywords}
cosmic rays -- gamma-rays: diffuse background -- gamma-rays: galaxies -- galaxies: starburst -- galaxies: star formation -- galaxies: ISM
\end{keywords}

%%%%%%%%%%%%%%%%%%%%%%%%%%%%%%%%%%%%%%%%%%%%%%%%%%

%%%%%%%%%%%%%%%%% BODY OF PAPER %%%%%%%%%%%%%%%%%%
%

\section{Introduction}

The extra-galactic $\gamma$-ray background (EGB) has been measured 
by \textit{EGRET}
~\citep{Sreekumar1998ApJ, Strong2004ApJ} and \textit{Fermi}-LAT
~\citep{Ackermann2015ApJ} 
to be a power-law of spectral index of around -2.3 from 100 MeV up to an exponential cut-off at 300 GeV, above which with detections persist to 820 GeV~\citep{Ackermann2015ApJ}. 
The EGB can be decomposed into a component arising from the $\gamma$-ray emission of \textit{resolved} extra-galactic sources, and a second component (sometimes referred to as the isotropic $\gamma$-ray background, IGRB) that emerges from the accumulation over redshift of all \textit{unresolved} $\gamma$-ray emitting sources beyond our Galaxy, extending to the furthest reaches of the observable Universe. 
 The physical origin of the unresolved component is thought to be a combination of
unresolved 
active galactic nuclei (AGN)
~\citep[e.g.][]{Inoue2011ApJ, Singal2012ApJ, Ajello2015ApJ} 
and star-forming galaxies (SFGs)
~\citep[e.g.][]{Bhattacharya2009RAA, Fields2010ApJ, Lamastra2017A&A}. A 
%point-like astrophysical sources, 
further \textit{cascade} component 
contributes as much as 50\% of the flux below a TeV~\citep{Coppi1997ApJ, Kneiske2008AA, Kalashev2009PhRvD, Berezinsky2011PhLB, Wang2011ApJ, Inoue2012PhRvD}, which arises from 
high-energy $\gamma$-rays undergoing 
pair-production and subsequent Compton scattering in the extra-galactic background light, EBL~\citep[e.g.][]{Madau1996ApJ}.

The exact balance between the sources of the EGB remains unsettled.
It has been argued that the majority of the flux originates in unresolved blazars~\citep{Inoue2009ApJ, Singal2012ApJ, Ajello2015ApJ}. These are complimented by radio galaxies~\citep{Inoue2011ApJ, DiMauro2014ApJ, Wang2016NatPh, Stecker2019arXiv}, also being active galactic nuclei (AGN) but discriminated from blazars by their viewing angle (sometimes referred to as misaligned AGNs, or MAGNs), and flat spectrum radio quasars, FSRQs~\citep{Ajello2012ApJ} -- see also~\citep[e.g.][]{Stecker1993ApJ, Abdo2010ApJ}.
Star-forming galaxies (SFGs) are also thought to make an important contribution, perhaps accounting for up to several tens of percent of the EGB intensity~\citep{Bhattacharya2009RAA, Fields2010ApJ, Tamborra2014JCAP, Lamastra2017A&A},\footnote{Other possible origins have also been considered, including contributions from galaxy clusters~\citep{Zandanel2015AA}, cascades from protons and/or heavy nuclei and their subsequent photo-disintegration/photo-pion production in cosmic radiation fields~\citep{Kalashev2009PhRvD, Ahlers2011PhRvD}, or
dark matter annihilation~\citep{Bertone2005PhR, Cirelli2011JCAP, Stecker2011ApJ, Bringmann2012PDU}. A more comprehensive discussion and assessment of candidate source populations is provided in the review paper~\citet{Fornasa2015PhR}.}, but their exact contribution remains poorly constrained~\citep[e.g.][]{Komis2019MNRAS}.
Intense starburst episodes experienced by SFGs 
yield abundant stellar end-products soon after the onset of star-formation.
The shocked violent astrophysical environments contributed by such end-products -- in particular supernova explosions and their remnants -- 
provide ample low-energy charged seed particles and magnetised shocks, which can boost the seeds to relativistic energies
(e.g. by Fermi acceleration -- ~\citealt{Fermi1949PhRv}) to form a rich interstellar reservoir of CRs.
Their containment within the internal environment of the host galaxy by a rapidly-amplified magnetic field~\citep{Schober2013A&A, Owen2018MNRAS}
particularly enhances the CR density in SFGs, which may undergo pion-producing hadronic interactions with the ambient gases of their host to drive $\gamma$-ray emission~\citep[e.g.][]{Pfrommer2017ApJ}.
Such emission has already been detected from several nearby starbursts,
 with several nearby SFGs having already been resolved in $\gamma$-ray observations with \textit{Fermi}-LAT -- 
M82 and NGC 253~\citep{Acero2009Sci, VERITAS2009Natur, Abdo2010ApJa, Rephaeli2015NucPhysBProc},
NGC 2146~\citep{Tang2014ApJ},
Arp 220~\citep{Peng2016ApJ, Griffin2016ApJ, Yoast-Hull2017MNRAS},
M33 and Arp 299~\citep{Xi2020arXiv, Ajello2020ApJ_SFG},
NGC 2403 and NGC 3424~\citep{Peng2019ApJ, Ajello2020ApJ_SFG},
and NGC 1068 and NGC 4945~\citep{Ackermann2012ApJa}
%\footnote{NGC 1068 and NGC 4945 host obscured AGN, which may also contribute to their $\gamma$-ray emission~\citep{Ackermann2012ApJa}.},
-- of these, some (M82 and NGC 253) also have higher energy counterpart detections with
{VERITAS} and {H.E.S.S.}~\citep{Acero2009Sci, Karlsson2009arXiv, Abdalla2018AA}. 

The superposition of sources in the EGB
must include contributions over a broad range of redshifts, particularly at lower energies (below around 30 GeV) where the cascade effect is less severe~\citep[e.g.][]{Gilmore2009MNRAS}. 
The cosmological propagation of $\gamma$-rays from their source must therefore be considered
when modelling the formation of the EGB.
At high energies, $\gamma$-ray photons interact with soft EBL photons and the cosmological microwave background (CMB) radiation
to produce a cascade of secondary charged leptons via pair-production~\citep{Heitler1954book, Madau1996ApJ}.
These secondary leptons can cool via Compton scattering (and, to a lesser degree, synchrotron emission in intergalactic magnetic fields)
to form the diffuse secondary flux of $\gamma$-rays~\citep[e.g.][]{Wang2011ApJ}.
Practically, this leads to an attenuation effect arising for higher energy $\gamma$-rays as they propagate, but an emission effect at lower energies along our line of sight to a source. The resulting energy spectrum can become heavily distorted over large distances -- particularly from sources located at higher redshifts, when intergalactic radiation fields would have been more intense than today and cascade losses more severe, and this must be carefully modelled.
This process is of often valuable to researchers, as the attenuation of high-energy $\gamma$-rays from distant sources can be used as a 
tool to probe and constrain EBL radiation fields which, in turn, can offer clues to the formation and evolution of galaxy populations over redshift~\citep[][]{Mazin2007AA, FermiLAT2018Sci}.
But in this work, we are concerned with the $\gamma$-rays themselves, and the modification and confluence of the energy spectra of a
broad distribution of sources located at different redshifts.
The attenuation of $\gamma$-rays is more severe at higher energies, and does not operate below $\sim$ 0.1 GeV (energies for which the $\gamma$-ray path length would presumably extend to distances comparable to the size of the visible Universe -- see, e.g.~\citealt{Neronov2012ApJ}) 

The Universe remains relatively transparent 
 to $z>5$ for $\gamma$-ray energies of up to around 30 GeV~\citep{Gilmore2009MNRAS}
but becomes optically thick by $z < 0.1$ in the TeV band~\citep{Franceschini2008AA, Dominguez2011MNRAS, Gilmore2012MNRAS, Inoue2013ApJ}.
This presents a challenge for spectral studies, where
higher energy emission would become inaccessible over relatively short cosmological distances.
However, the CR spectrum of SFGs peaks at around 1-10 GeV and, given the weak energy-dependence of the pp-interaction cross section above its interaction threshold~\citep{Kafexhiu2014}, their $\gamma$-ray emission would presumably reflect this and much of their flux would fall below the energies most strongly affected by EBL attenuation.
The peak of cosmic star-formation (and presumably the redshift at which SFGs are most abundant)
arises at approximately $z\approx2$~\citep{Madau2014ARAA, FermiLAT2018Sci} -- a distance from which EBL transmittance at 10s of GeV is reasonable. 
We therefore expect that SFGs would make a substantial contribution to the observed EGB at these energies. 

Studies of the $\gamma$-ray background with \textit{Fermi}-LAT
have recently revealed small-scale anisotropies between 0.5 and 500 GeV~\citep{Fornasa2016PhRvD, Ackermann2018PhRvL}, thought to arise from two distinct source populations (SFGs and BL Lac. objects).
In this work, we investigate such signatures imprinted by a SFG contribution to the $\gamma$-ray background and assess their form and its sensitivity to the underlying source population's basic characteristics.
We consider a model for the emission, attenuation, reprocessing and cosmological propagation of $\gamma$-rays, also accounting for 
the redshift-evolution of their star-forming galaxy populations.
We consider the characteristic separation of galaxies at a given redshift, described by the galaxy power spectrum~\citep{Tegmark2004PhRvD},
and compute how the resulting redshift-integrated length-scale would become imprinted into the $\gamma$-ray background.
We demonstrate how the resulting angular power spectrum of the anisotropies in the EGB is sensitive to the properties and redshift evolution of the source galaxy population, and consider the observational prospects of such a signature with ongoing \textit{Fermi}-LAT observations, and the up-coming Cherenkov Telescope Array (CTA).

This paper is organised as follows. In Section~\ref{sec:section2}, we outline the CR interactions and discuss the $\gamma$-ray emission from SFGs. We consider the relevant pair-production processes arising within a galactic environment, and assess the impacts these have on the emitted $\gamma$-ray spectrum from a SFG. In Section~\ref{sec:section3}, we consider the properties of SFG populations, model the cosmological propagation of $\gamma$-rays and introduce our EGB anisotropy model. We present our results in section~\ref{sec:section4} and show the relation between EGB anisotropy signatures and key parameters in our model. We also consider the detection prospects of such signatures. We provide a brief summary and draw conclusions in Section~\ref{sec:section5}.

\section{$\gamma$-ray emission from star-forming galaxies}
\label{sec:section2}

We adopt the following notation convention. We express particle energies in terms of their Lorentz factor, e.g. for protons, the total energy $E_{\rm p} = \gamma_{\rm p} \;\!m_{\rm p}c^2$, which is related to the proton kinetic energy by $T_{\rm p} = (\gamma_{\rm p}-1) m_{\rm p} c^2$. Photon energies (including $\gamma$-rays) are expressed in terms of the dimensionless quantity $\epsilon_{\gamma} = E_{\gamma}/m_{\rm e}c^2$, i.e. normalised to the electron rest mass energy. 
%Where appropriate, differential fluxes and spectra are plotted, e.g. $E_{\gamma} \;\! {\rm d}n_{\gamma}/{\rm d}E_{\gamma}$.

\subsection{Cosmic ray interactions}

Energetic hadronic cosmic rays may interact through various channels, 
however
the internal conditions of typical star-forming galaxies would strongly favour proton-proton (hereafter pp) pion-production mechanisms~\citep{Owen2018MNRAS, Owen2019AA}.
%The interstellar medium of s is because of the abundance of rich, dense pockets of gas fuelling  star-formation which provides an important target for the energetic protons. So-called p-$\gamma$ 
Hadronic interactions with radiation fields (photo-pion or photo-pair production) are comparatively unimportant, despite of the intense radiation fields generated by the young stellar populations~\citep{Owen2018MNRAS}. % which do not attain sufficient energy density to compete with pp processes.
The pp interaction can arise above a threshold proton kinetic energy of $T_{\rm p}^{\rm th}/c^2 = 2m_{\pi^0} + m_{\pi^0}^2/2m_{\rm p} = 0.28~\text{GeV}/c^2$, and involves a two-stage mechanism: the first step is the formation of resonance baryons via $\rm{p}\rm{p} \rightarrow \rm{p}  \Delta^{+~\;}$ or $\rm{p}\rm{p} \rightarrow \rm{n} \Delta^{++}$~\citep{Almeida1968PR, Skorodko2008EPJA}, while the second step is their decay (on timescales of $5.63\times 10^{-24}$ s -- see~\citealt{Patrignani2016ChPh}) according to
\begin{align}
\label{eq:pp_interaction_decay1}%
	\Delta^{+~\;} \rightarrow\begin{cases}%
				 \rm{p} \pi^{0}  \xi_{0}(\pi^{0}) \xi_{\pm}(\pi^{+} \pi^{-}) \\[0.5ex]%
				 \rm{p}  \pi^{+}  \pi^{-}  \xi_{0}(\pi^{0}) \xi_{\pm}(\pi^{+} \pi^{-}) \\[0.5ex]%
				 \rm{n}  \pi^{+}  \xi_{0}(\pi^{0}) \xi_{\pm}(\pi^{+} \pi^{-})\\[0.5ex]%
				 \end{cases} \ ,
\end{align}
or
\begin{align}
\label{eq:pp_interaction_decay2}%
	\Delta^{++} \rightarrow\begin{cases}%
				 \rm{p} \pi^{+} \xi_{0}(\pi^{0}) \xi_{\pm}(\pi^{+} \pi^{-}) \\[0.5ex]%
				 \rm{n} 2\pi^{+} \xi_{0}(\pi^{0}) \xi_{\pm}(\pi^{+} \pi^{-})\\[0.5ex]%
			\end{cases} \ ,
\end{align}%
    for which the terms $\xi_{0}$ and $\xi_{\pm}$ are the energy-dependent multiplicities of the three pion species \citep{Jain1993physrep, Lebiedowicz2014IFJPAN}.
  Despite the energy-dependence of the multiplicities
  \citep{Almeida1968PR, Blattnig2000, Skorodko2008EPJA},
the overall intrinsic production rate of each pion species is relatively energy-independent, with ratios of $\{\pi^+, \pi^-, \pi^0\} = \{ 0.6, 0.1, 0.3\}$ 
  arising at 1~GeV, and slowly varying to
  $\{ 0.3, 0.4, 0.3\}$ by 50~GeV, with negligible evolution thereafter~\citep{Jacobsen2015}.
  $\gamma$-ray production proceeds (with a branching ratio of 98.8\%) through neutral pion decays 
  $\pi^0 \rightarrow 2\gamma$
   on a timescale of $8.5 \times 10^{-17}\;\!{\rm s}$ \citep{Tanabashi2018PRD} and, given the weak energy-dependence of the pp interaction cross-section and the $\pi^0$ formation multiplicity, would yield a $\gamma$-ray spectrum closely tracing the shape of the proton spectrum driving the emission. $\gamma$-ray emission can also arise from inverse Compton scattering of secondary electrons injected by charged pion decays. However, the resulting emissivity by this channel is not expected to dominate at energies above 0.1-1 GeV~\citep[e.g.][]{Chakraborty2013ApJ, Pfrommer2017ApJ} and so is not included.
   
   \subsubsection{Hadronic interaction rate}
   
    The volumetric rate at which pp interactions arise is given by
    \begin{equation}
        \label{eq:rate_pp}
        \dot{n}_{\rm p\pi}(\gamma_{\rm p}) = \langle n_{\rm H}\rangle \;\! n_{\rm p}(\gamma_{\rm p}) \;\!{c}\;\!\sigma_{\rm p\pi}(\gamma_{\rm p}) \ , 
    \end{equation}
where $\langle n_{\rm H}\rangle$ is the average ambient gas density
%\footnote{The inhomogeneity of the medium is relatively inconsequential: the magnetic field structure in dense molecular clouds and clumps could cause some containment and elevated interaction rate in their cores, but this is only within a factor of a few~\citep{Owen2020MNRAS} -- far less than other inherent uncertainties and approximations in our model. Otherwise, there is no indication that the CR interaction rate and/or $\gamma$-ray emissivity would differ in over-dense or under-dense regions beyond a proportional scaling with their gas density, meaning the mean ISM density treatment adopted here is sufficient to estimate the $\gamma$-ray emissivity of the entire host galaxy.} 
within the host galaxy, $n_{\rm p}$ is the CR proton density, 
%$\xi$ is introduced as a parameter to account for the clumpiness of the interstellar gas in the host galaxy (see section~\ref{sec:multiphase_ism}), 
and $\sigma_{\rm p\pi}$ is the total inelastic pp interaction cross section, which may be parameterised as
\begin{equation}
\label{eq:pp_cs}
   \sigma_{\rm p\pi} = \left( 30.7 - 0.96\ln(\chi) + 0.18(\ln\chi)^{2} \right)\left( 1 - \chi^{-1.9}   \right)^{3}~\rm{mb} \ ,
\end{equation}
   \citep{Kafexhiu2014}, where $\chi = T_{\rm p}/T_{\rm p}^{\rm th}$, for $T_{\rm p}^{\rm th} = (\gamma_{\rm p}^{\rm{th}}-1)\;\!m_{\rm p}c^2$ is the threshold proton kinetic energy, below which the interaction cannot occur. Equation~\ref{eq:pp_cs} therefore represents the volumetric loss rate of CR protons due to the pp process within the interstellar medium (ISM) of the host starburst galaxy.
   This is different (although related) to the production rate of $\gamma$-rays, which relies on the formation and subsequent decay of neutral pions. The differential $\gamma$-ray inclusive cross section by the ${\rm pp} \rightarrow {\rm pp} \pi^0$ interaction channel may be written as
   \begin{equation}
       \frac{{\rm d}{\sigma}_{\rm p\gamma}(\gamma_{\rm p}, \epsilon_{\gamma})}{{\rm d}\epsilon_{\gamma}} = \mathcal{P}(\gamma_{\rm p})\times\mathcal{F}(\gamma_{\rm p}, \epsilon_{\gamma})  \ ,
   \end{equation}
   where the peak function $\mathcal{P}$ and spectral form $\mathcal{F}$ are also well-parametrised to an accuracy of better than 10 per cent by~\citet{Kafexhiu2014}.
   
\subsubsection{$\gamma$-ray production in starburst galaxies}
\label{sec:gamma_ray_from_gal}

We compute the $\gamma$-ray spectral emissivity from the CR proton density, $n_{\rm p}(\gamma_{\rm p})$ (see section~\ref{sec:cr_flux_normalisation}; note that this is also a differential quantity such that $n_{\rm p}(\gamma_{\rm p})\;\!{\rm d}\gamma_{\rm p}$ is the number density of CR protons in the energy interval ${\rm d}\gamma_{\rm p}$), and the inclusive differential $\gamma$-ray production cross-section.\footnote{Other works~\citep[e.g.][]{Peretti2020MNRAS} instead consider the $\gamma$-ray flux density of M82 as a prototype, and model the $\gamma$-ray emission of other SFGs by scaling this with star-formation rate.}
The spectral emissivity of $\gamma$-rays may be written as
\begin{equation}
\frac{{\rm d}\dot{n}_{\gamma}(\epsilon_{\gamma})}{{\rm d}\epsilon_{\gamma}} = 
c\;\! \langle n_{\rm H}\rangle \;\! \int_{\gamma_{\rm p}^{\rm th}}^{\gamma_{\rm p}^{\star}} \frac{{\rm d}{\sigma}_{\rm p\gamma}(\gamma_{\rm p}, \epsilon_{\gamma})}{{\rm d}\epsilon_{\gamma}}\;\! n_{\rm p}(\gamma_{\rm p}) \;\! {\rm d}\gamma_{\rm p} \ ,
\label{eq:emiss_spec}
\end{equation}
where $c$ is the speed of light, and where we set $\gamma_{\rm p}^{\star} = 50~\text{PeV}/m_{\rm p}c^2$ as the upper limit for the acceleration of CR protons in starbursts~\citep[as argued in][]{Peretti2019MNRAS}.

\subsection{Cosmic ray spectrum and energy budget}
\label{sec:cr_flux_normalisation}

To compute the $\gamma$-ray spectral emissivity of a galaxy (equation~\ref{eq:emiss_spec}), we require knowledge of the internal CR spectrum. 
Typically, this is well-described by a simple power-law, with a spectral index $\Gamma$ between -2.1 and -2.7 depending on the exact environment~\citep{Kotera2011ARAA}. In star-forming regions, the particle spectrum is presumably freshly accelerated and would be described by a less-steep spectral index, with more CRs at higher energies. Here, we initially adopt a proton spectrum of $\Gamma = -2.1$, being similar to 
the characteristic value (of between -1.9 and -2.3) inferred for local SFGs detected in $\gamma$-rays~\citep[see, e.g.][]{Tamborra2014JCAP, RojasBravo2016MNRAS, Ajello2020ApJ_SFG}.\footnote{While spectral indices are observed for the $\gamma$-ray emission from nearby SFGs, it is expected that the $\gamma$-ray spectrum above $\sim$ 1 GeV would closely follow the 
CR proton spectrum, because of 
the limited energy-dependence of the inclusive $\pi^0$ formation cross section -- e.g.~\cite{Kafexhiu2014}.} This choice of index is also comparable to regions within the Milky Way where CRs are thought to be freshly accelerated, i.e. towards the galactic ridge~\citep[see, e.g.][]{Allard2007APh, Kotera2010JCAP}.
We relax this choice later, in section~\ref{sec:section4}, where we consider alternative CR index values within this range.

\subsubsection{Cosmic ray luminosity}

We estimate the CR proton density within a SFG using equation~\ref{eq:final_protons_eq} (see Appendix~\ref{sec:cr_proton_density} for details), where we consider that the CR proton spectrum and density can be parameterised by just four quantities: the star-formation rate of the host galaxy, $\mathcal{R}_{\rm SF}$, the size of the nuclear starburst region, $R$, the CR spectral index, $\Gamma$, and the maximum CR proton energy, $E_{\rm max}$. We argue that there is insufficient motivation to consider substantial variation of other quantities, such as those pertaining to the CR diffusion coefficient (equation~\ref{eq:parametric_diffusion_coeff}), ISM density/structure, 
or the characteristic fraction of CRs advected by galactic outflows,
and we fix these are their values stated in Appendix~\ref{sec:cr_proton_density}. 
%Further investigation of any variation of these quantities relating to the detailed substructure and configuration of SFG environments remains beyond the scope of this study and is left to future work. 

%$L_{\rm CR, eff}$ is the total power in the CR protons, which can be expressed in terms of SN event rate ${\cal R}_{\rm SN}$. While the CR density profile may be found by calculating equation~\ref{eq:proton_solution} via a Monte-Carlo method, previous works~\citep{Owen2018MNRAS, Owen2019AA} found the radial profile to be relatively constant through the host galaxy.
%Given that the  flux is not of interest in the present work, we argue that a uniform approximation can be adopted to avoid the need for detailed numerical modelling of the internal variation of $n_{\rm p}$ within the host system. As such, we  (i.e. constant), and we simplify equations~\ref{eq:proton_solution} and~\ref{eq:proton_atten} as
%\begin{equation}
%n_{\rm p} = \; \frac{35 R^2 \;\! \mathcal{L}_{0} \;\! \mathcal{A}(\zeta_{\rm p\pi})}{108 \;\! D(E_{\rm p})} \;\! \frac{\partial}{\partial E_{\rm p}} \left(\frac{E}{E_0}\right)^{-\Gamma} \biggr\vert_{E_{\rm p}} \ ,
%\end{equation}

\subsubsection{Internal attenuation of $\gamma$-rays}
\label{sec:internal_attenuation}

\begin{figure}
    \centering
    \includegraphics[width=\columnwidth]{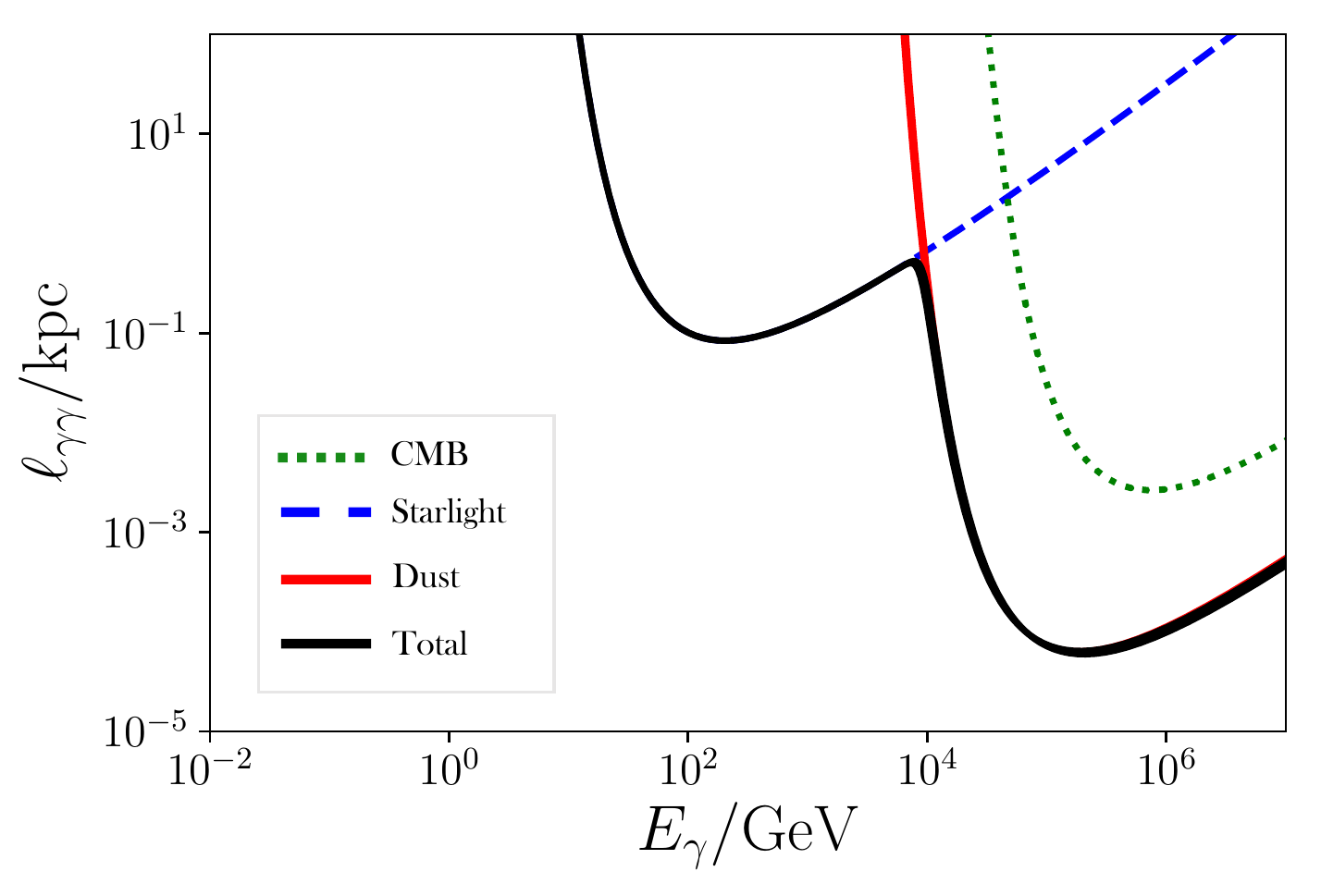}
    \includegraphics[width=\columnwidth]{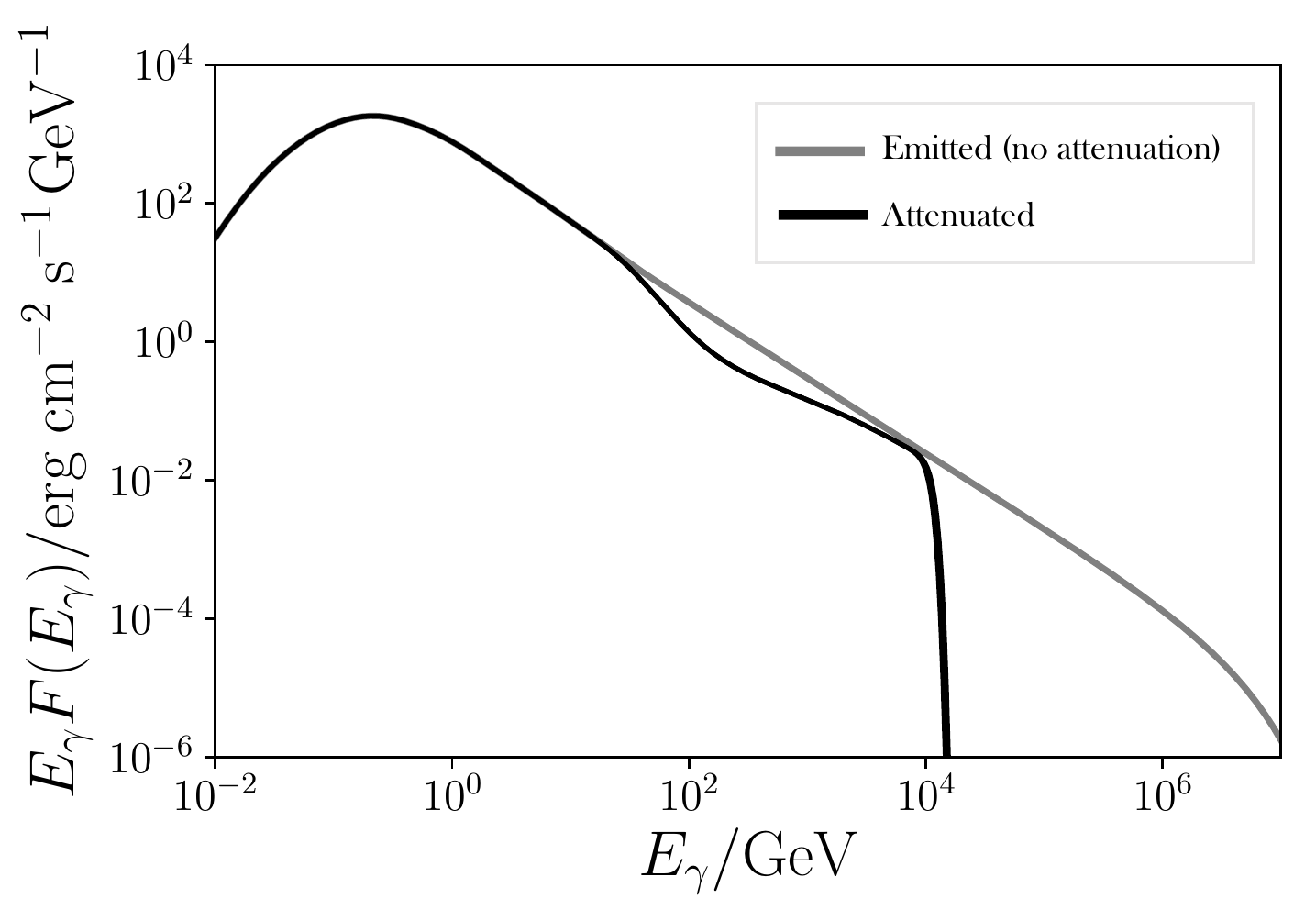}
    \caption{{\bf Top}: Effective path lengths of $\gamma$-rays within a SFG nucleus. Attenuation due to the CMB is most effective at lower energies, but is strongly dominated by dust at around 10 TeV. This is for the fiducial model galaxy of $\mathcal{R}_{\rm SF} = 10~\text{M}_{\odot}\;\!\text{yr}^{-1}$, $R = 0.1~\text{kpc}$ and at $z=2$. 
    {\bf Bottom}: $\gamma$-ray emission arising from the pp interaction (and subsequent $\pi^0$ decay) for the fiducial model is shown in grey. The black line shows the effective $\gamma$-ray emission spectrum from the galaxy, accounting for the $\gamma$-ray absorption within the SFG nucleus.
    The uncertainties in the dust temperature propagate to the shaded region (black) in both plots.}
    \label{fig:spec_and_attenuation}
    \end{figure}

While the production of high-energy $\gamma$-rays within SFGs is predominantly regulated by the hadronic pp interactions of CRs (see Figure~\ref{fig:spec_and_attenuation}), their resulting $\gamma$-ray emission spectra is more complicated than this would imply. $\gamma$-ray absorption processes would operate within the internal environment of a SFG, substantially modifying the ensuing emitted spectrum. Recent works (e.g. \citealt{Vereecken2020arXiv}) have considered that sufficiently dense gas clouds within the ISM of a host galaxy could significantly attenuate $\gamma$-rays through pair-production on the dense gas. This is compelling, as such dense clouds would act as a CR beam dump via the pp-interaction, making these the principal sites of $\gamma$-ray production by this channel.
However, such a mechanism across a galaxy would require very substantial gas column densities along many ISM lines of sight. This would imply a heavy loading of the ISM with large, dense clouds. While this cannot be ruled-out -- and may be one of several processes operating to modify the $\gamma$-ray spectrum of a galaxy on a global scale -- in this work we instead consider the $\gamma$-ray attenuation that would result from ambient radiation fields (in particular, those due to the stellar population and the component of the stellar radiation reprocessed to infra-red, IR wavelengths by interstellar dust), which we find would dominate $\gamma$-ray attenuation under averaged ISM conditions in SFGs (see Appendix~\ref{sec:atten_comp} for details).
The relative importance of possible $\gamma$-ray attenuation processes, and their dependence on the internal interstellar environment and multi-phase structure of SFGs, is left to future work.

$\gamma\gamma$ pair-production between high-energy $\gamma$-rays and a low energy target photons provided by the CMB, starlight or dust-reprocessed starlight, proceeds as:
\begin{equation}
   \gamma + \gamma \rightarrow e^+ + e^- \ ,
\end{equation}
at a rate of
\begin{equation}
    \dot{N}_{\gamma\gamma}(\epsilon_{\gamma}) = c\;\!\int_0^{\infty} {\rm d}\epsilon \;\! n_{\rm ph}(\epsilon)\;\!\sigma_{\rm \gamma\gamma}(\epsilon_{\rm r}) \ ,
    \label{eq:rate_gg}
\end{equation}
where $n_{\rm ph}$ is the spectral number density of target photons, $\sigma_{\rm \gamma\gamma}$ is the $\gamma\gamma$ interaction cross section~\citep[see, e.g.][]{Gould2005_book} and $\epsilon_{\rm r} \approx \epsilon \epsilon_{\gamma}/2$ is the invariant interaction energy for an isotropic radiation field. The electron pairs formed in this process can predominantly cool by Compton up-scattering photons in ambient radiation fields, or thermalise in the interstellar gas depending on their energy (with other processes arising at a lower rate -- see~\citealt{Owen2018MNRAS} for a comparison of various cooling timescales experienced by electrons in typical SFGs).
Electrons below $\sim 100~{\rm MeV}$ predominantly thermalise in less than 1 Myr in ISM conditions~\citep{Owen2018MNRAS}, corresponding to a diffusive length-scale of $\sim$ 1 kpc. 
However, most electrons are injected at much higher energies than this, above 10s of GeV (reflecting the energies where $\gamma$-ray attenuation is strongest -- see Figure~\ref{fig:spec_and_attenuation}).
For these, thermalisation timescales are much longer so electromagnetic {cascades} tend to develop instead, where electrons Compton up-scatter ambient interstellar radiation field (ISRF) photons to high-energies (typically to form so-called secondary $\gamma$-rays or X-rays; see, e.g.~\citealt[][]{Chakraborty2013ApJ}). At $100~{\rm GeV}$, for instance, electrons would thermalise over $\sim$ 1 Gyr in the typical SFG environment considered in~\citealt{Owen2018MNRAS}, while their Compton scattering timescale would be just a few kyr. The up-scattered secondary photons may undergo further pair-production, if they are of sufficient energy.\footnote{If each of the secondary electrons adopts half of the energy of the primary $\gamma$-ray, the resulting Compton-scattered secondary $\gamma$-rays would have a peak energy of $E_{\gamma, 2} \approx \left(E_{\gamma, 1}/m_{\rm e} c^2\right)^2 \;\! E^{\rm peak}_{\rm ph}$ where $E^{\rm peak}_{\rm ph}$ is the peak energy of the target radiation field, and $E_{\gamma, i}$ for $i=\{ 1, 2 \}$ are the primary and peak secondary $\gamma$-ray energies respectively. For a 100 GeV primary $\gamma$-ray, the characteristic secondary energy would be of order $\sim$ 10 MeV, while for a 100 TeV primary, it would be of order $\sim$ 10 TeV.}  Alternatively they may escape from the host galaxy, modifying the emitted $\gamma$-ray spectrum from the SFG. However, given that the majority of the secondary $\gamma$-ray emission develops from the attenuation of the highest energy primary $\gamma$-rays (cf. Figure~\ref{fig:spec_and_attenuation}), for which the flux is lowest (due to the power-law nature of the emissivity), their contribution to the emitted spectrum would be  negligible~\citep{Fitoussi2017MNRAS}.
%\footnote{We later show (in section~\ref{sec:cosmological_gamma_propagation}) that this assumption is not valid for intergalactic conditions, where the EBL attenuation is also important for intermediate energy $\gamma$-rays, for which the flux (and subsequent secondary $\gamma$-ray production) is much higher. Within an ISRF, the path length of $\gamma$-rays between 10s of GeV and 10s of TeV is typically large enough for them to escape from their kpc-scale host galaxy without substantial attenuation, but their path lengths in the EBL (which is comprised of components with similar black-body temperatures to the ISRF) leads to strong attenuation over much greater intergalactic/cosmological distances.} 
As such, we argue that the final emitted $\gamma$-ray spectrum of a SFG can be well-described by the $\gamma$-ray emissivity from hadronic interactions, modified by their attenuation through pair production. The negligible secondary {cascade} emission from within the ISM of the host galaxy is not included in our model.

Without loss of generality, we define the characteristic $\gamma$-ray path length in a radiation field as $\ell_{\gamma\gamma}(\epsilon_{\rm \gamma}, x) = c/\dot{N}_{\gamma\gamma}(\epsilon_{\rm \gamma}, x)$. This is the distance over which an interaction would typically arise under conditions specified at location $x$.
It can be associated with a pair-production $\gamma$-ray optical thickness by
\begin{equation}
   \tau_{\gamma\gamma}(\epsilon_{\rm \gamma}, x) = \int_0^x \;\! {\rm d}x' \;\!\ell_{\rm \gamma \gamma}^{-1}(\epsilon_{\rm \gamma}, x') \ ,
   \label{eq:tau_pair_production}
\end{equation}
over some path length $x$. In an isotropic black-body radiation field, 
this may be expressed as
%\footnote{This follows from the reduction of equation~\ref{eq:rate_gg} for a black-body radiation field~\citep[see][for details]{Gould1967PhRv, Brown1973ApL}.}
\begin{equation}
    \tau_{\gamma\gamma}^{\rm bb}(\epsilon_{\rm \gamma}, x) = \frac{2 \alpha_{\rm f}^2}{\lambda_{\rm C}}\;\! \int_0^x\;\!\Theta^3(x')\;\!\mathcal{J}(\epsilon_{\rm \gamma}, x')\;\!{\rm d}x' \ ,
    \label{eq:tau_gg}
\end{equation}
\citep{Gould1967PhRv, Brown1973ApL, Dermer2009_book}. Here, $\alpha_{\rm f}$ is introduced as the fine structure constant, $\lambda_{\rm C}$ is the electron Compton wavelength, $\Theta(x) = k_{\rm B} T(x)/m_{\rm e}c^2$, with $m_{\rm e}$ as the electron rest mass,
$c$ as the speed of light and
$k_{\rm B}$ is the Boltzmann constant,
 defines the dimensionless temperature at some position $x'$, and the function $\mathcal{J}(...)$ is given by:
\begin{equation}
    \mathcal{J}(\epsilon_{\rm \gamma}, x) = \frac{1}{\epsilon_{\gamma}^2\Theta^2(x)}\;\!\int_{1/\epsilon_{\gamma}\Theta(x)}^{\infty}\;\!\frac{\varphi\left( y\;\!\epsilon_{\gamma}\;\! \Theta(x) \right)\;\!{\rm d}y}{\exp(y) -1}
\end{equation}
\citep[e.g.][]{Zdziarski1989ApJ}, where the term
\begin{equation}
    \varphi(\epsilon^{\star}) = \frac{2}{\pi r_e^2}\int_1^{\epsilon^{\star}}\;\!{\rm d}\epsilon_{\rm r}\;\!\epsilon_{\rm r}\;\!\sigma_{\gamma\gamma}(\epsilon_{\rm r})
    \label{eq:varphi_def}
\end{equation}
specifies the change in the scattering cross-section compared to the (classical) Thomson cross-section $\sigma_{\rm T} = \pi r_e^2$. $\epsilon^{\star} = \epsilon \epsilon_{\gamma}$ is the combined photon energy, and $r_e$ is the classical electron radius. This is evaluated in~\citealt{Gould1967PhRv}~(see also~\citealt{Brown1973ApL}).
Equation~\ref{eq:tau_gg} can be used to quantify the $\gamma$-ray attenuation factor within the host galaxy. Along a single line of sight $s$, this would simply be
\begin{equation}
    \mathcal{A}(\epsilon_{\rm \gamma}, s) = \exp\left\{ -\tau_{\gamma\gamma}^{\rm bb}(s, \epsilon_{\gamma}) \right\} \ ,
\end{equation}
however, when averaged through an extended SFG source (modelled as a uniformly attenuating sphere of radius $R$), we instead adopt an approximate characteristic attenuation factor specified by the size of the nucleus $R$ and the effective path length of the $\gamma$-rays, $\ell_{\gamma\gamma}$:
\begin{equation}
\mathcal{A}(\zeta) = \exp\left(-{\zeta^2}\right) \ ,
\label{eq:final_mean_atten_maintext}
\end{equation}
(see Appendix~\ref{sec:appendix_a} for details) where $\zeta = (R/\ell_{\gamma\gamma})^{1/2} = \tau_{\rm bb}^{1/2}(\epsilon_{\gamma}, R)$.

We consider that $\gamma$-ray attenuation within a SFG is dominated by three radiation fields: (1) the CMB; (2) the ISRF from stars, and (3) the re-processed ISRF by interstellar dust. These may each be described by a Planck function of the form
\begin{equation}
n_{\rm ph}^{\rm bb}(\epsilon; \Theta) = \frac{8 \pi f_{\rm ph}}{\lambda_{\rm C}^3} \frac{\epsilon^2}{\exp\left(\epsilon/\Theta\right) - 1} \ ,
\end{equation}
where $\Theta = k_{\rm B}T/m_{\rm e}c^2$ is the dimensionless temperature of the radiation field, and $f_{\rm ph}$ is the dilution factor for geometrically distributed sources. The CMB is an undiluted radiation field, so $f^{\rm CMB}_{\rm ph} = 1$. Its temperature is a function of redshift, and is described by $T^{\rm CMB}(z) = T_0(1+z)$, where $T_0 = 2.73~\text{K}$~\citep{Planck2018} is the temperature of the CMB today; as a baseline model in the following results and calculations, we adopt a redshift of $z=2$ (unless specified otherwise). 
The ISRF components due to stars and re-processed emission by dust would be a diluted black-body, as the emission originates from the stars. The dilution factor in these two cases can be determined from the photon density in each radiation field. 
In general, this may be estimated as
\begin{equation}
f_{\rm ph}
 \approx \frac{L \;\! \lambda_{\rm C}^3}
 {96 \pi^2 R^2 m_{\rm e} c^3 \Theta^4 \Gamma(3) \zeta(3)} \ ,
 \label{eq:fraction_radiation_stars_dust}
\end{equation}
which is the ratio of the estimated photon density from the diluted black-body radiation field, compared to that expected for an undiluted black-body of the same temperature.
Here, $L$ is the total luminosity of the sources, $\Gamma(...)$ is the gamma function, and $\zeta(...)$ is the Riemann zeta function. As a baseline choice, we adopt a characteristic value of $R = 0.1~{\rm kpc}$ for the size of a SFG nucleus, being comparable to the that of nearby starbursts, for example NGC 253~\citep{Weaver2002ApJ}, for which $R\approx 0.1~{\rm kpc}$, or M82 (found to have a core of $0.25~{\rm kpc}$ -- see \citealt{deGrijs2001}).

Young stars dominate the radiative emission from the stellar population of a SFG, and the dust optical depths are so great that a large fraction of the bolometric SFG luminosity is re-processed and re-radiated to IR wavelengths~\citep{Kennicutt1998ApJ}. We consider that the total dust-reprocessed luminosity is comparable to the luminosity integrated over the full mid and far IR spectrum (8-1000$\mu$m). For SFGs, most of this emission will fall in the 10-120 $\mu$m spectral band~\citep{Kennicutt1998ARAA}. As such,
the luminosity of the dust emission from a SFG, $L = L^{\rm dust}_{\rm IR}$, is strongly coupled to its $\mathcal{R}_{\rm SF}$,
%\footnote{Young stars dominate the radiative emission from the stellar population of a SFG, and the dust optical depths are so great that a large fraction of the bolometric SFG luminosity is re-processed and re-radiated to IR wavelengths~\citep{Kennicutt1998ApJ}. We consider that the total dust-reprocessed luminosity is comparable to the luminosity integrated over the full mid and far IR spectrum (8-1000$\mu$m). For SFGs, most of this emission will fall in the 10-120 $\mu$m spectral band~\citep{Kennicutt1998ARAA}.}
via:
\begin{equation}
L^{\rm dust}_{\rm IR} = 2.2\times 10^{43} \;\!\left(\frac{\mathcal{R}_{\rm SF}}{1\;\!{\rm M}_{\odot}\;\!{\rm yr}^{-1}}\right)~{\rm erg}\;\!{\rm s}^{-1}
\label{eq:dust_conversion}
\end{equation}
~\citep{Kennicutt1998ApJ}, which is derived by applying the models of~\citep{Leitherer1995ApJS} for continuous starburst episodes of age 10-100 Myr, and a \citet{Salpeter1955ApJ} initial stellar mass function between 0.1 and 30 ${\rm M}_{\odot}$. Presumably, this scaling relation would not be strongly sensitive to the exact choice of lower or upper IMF mass cut-off, if less than $\sim$ 1 ${\rm M}_{\odot}$ or above 30 ${\rm M}_{\odot}$, 
for which the luminosity or number of stars (respectively) would not be substantial.
We adopt this relation, which holds for the vast majority of SFGs~\citep{Bergvall2016AA}, where the star-forming burst durations do not greatly exceed 100 Myr~\citep{Kennicutt1998ARAA}. 
Interstellar dust emission is typically dominated by that from large grains ($>0.01 \mu{\rm m}$), which are in thermal equilibrium with ambient interstellar radiation~\citep[e.g.][]{Desert1990AA}.
The temperature of their emission $T_{\rm dust}$ (encoded by $\Theta = \Theta_{\rm dust} = k_{\rm B}T_{\rm dust}/m_{\rm e}c^2$) in SFGs typically takes a characteristic value of a few tens of K. There is evidence for some redshift evolution~\citep{Magdis2012ApJ, Magnelli2014AA, Bethermin2015AA}, with effective temperatures increasing from typical values of around 25 K at $z=0$, to around 40 K by $z=4$~\citep[e.g.][]{Schreiber2017AA, Schreiber2018AA}. We adopt the empirical power law of~\citet{Schreiber2018AA} to model this, 
\begin{equation}
T_{\rm dust} = \left[32.9 \pm 2.4 + (z-2) (4.60 \pm 0.35) \right]~{\rm K} \ ,
\label{eq:dust_relation}
\end{equation}
where uncertainties are propagated through our model.
Our fiducial model considers a redshift of $z=2$, near the peak of cosmic star-formation~\citep{Madau2014ARAA}. This gives a corresponding dust temperature of $32.9\pm2.4$ K.
%We adopt a fiducial value of $T_{\rm dust} = 30 {\rm K}$, roughly corresponding to a characteristic value at $z=2$, but explore the impact of variation and redshift evolution of this parameter in section~\ref{sec:dust_evo}.
We model the dust-reprocessed radiation field to be spatially homogeneous and isotropic within the host SFG nucleus (up to a radius of $R$). The impact of detailed interstellar variations of the dust emission within SFGs (e.g. the clumpy distributions found in~\citealt{Bassett2017MNRAS}) is left to future work.

The total stellar radiative output power of stars in an SFG, $L = L^{\star}$, is dominated by young, massive, O and B type stars. It can be estimated from the dust luminosity $L^{\rm dust}_{\rm IR}$, using:
\begin{equation}
L^{\star} = \frac{(1-\eta) L^{\rm dust}_{\rm IR}}{0.4 - 0.2 f_{\rm abs} + 0.6 \beta} \ ,
\end{equation}
\citep{Inoue2000PASJ},
where
$f_{\rm abs} = 0.26$ is the fraction of ionising stellar photons absorbed by ISM Hydrogen~\citep[from][which derives the value from the Orion nebula]{Petrosian1972ApJ},
$\beta = 0.6$ is the averaged dust-absorption efficiency of non-ionising photons from central sources in ionised, star-forming regions~\citep[from][which uses the extinction curve of the Galaxy]{SavageARAA1979}, and
$\eta = 0.5$ is the fraction of IR emission attributed to diffuse ISM gas, being distinct from the emission from star-forming regions~\citep{Helou1986ApJ}.
This approach is valid for both strong starbursts (which emit almost all of their energy in IR -- see~\citealt{Soifer1987ARAA}) as well as moderate starbursts (where a large fraction of the stellar radiation may not be reprocessed by dust -- see~\citealt{Buat1996AA}). We set the temperature of this radiation field to be $T^{\star} = 3\times10^4~{\rm K}$, to reflect the temperature of the dominant source population of massive O and B type stars.

The attenuative effects of these radiation fields on the emitted spectrum are demonstrated in Figure~\ref{fig:spec_and_attenuation} for our fiducial case ($\mathcal{R}_{\rm SF} = 10~\text{M}_{\odot}\;\!\text{yr}^{-1}$, $R = 0.1~\text{kpc}$ and at $z=2$). This demonstrates the severe impact of interstellar dust, which completely attenuates $\gamma$-rays above 10 TeV in this case. The CMB has some impact on lower-energy $\gamma$-rays, and would become more severely attenuating at higher redshifts. The (un-processed) starlight is comparatively unimportant, with a large fraction of the stellar emission having been reprocessed to IR wavelengths by interstellar dust.

%Key references:
%On the properties (internal) of star-forming galaxies, and how this will be needed to inform and parameterise the gamma-ray production and propagation through these environments, see~\citep[from JDS discussions, see][]{Silverman2018ApJ, Rujopakarn2019arXiv, Silverman2018ApJ, Kashino2013ApJ, Kashino2017ApJ}

%Gamma-ray emission modelling for starburst galaxies (and propagation of cosmic rays within their ISM) -- see work by the Ahronian group so far...

%~\citep{2019arXiv191106163P, Peretti2019MNRAS}. Several aspects of their approach may be applicable here too.

%Without loss of generality, we express the intrinsic $\gamma$-ray spectral luminosity of a star-forming galaxy in terms of the spectral emissivity as:
%
%\begin{equation}
%   \frac{{\rm d}L_{\gamma}^{\rm g}(E_{\gamma}, ...)}{{\rm d}E_{\gamma}} = 4\pi\;\!\mathcal{G}\;\!m_{\rm e}c^2\;\!\epsilon^2 \;\!\mathcal{A}^{\rm ISM}(\epsilon)\int_{r=0}^{R}\;\!\frac{{\rm d}\dot{n}_{\gamma}(\epsilon, r, ...)}{{\rm d}\epsilon}\;\!r^2\;\!{\rm d}r \ ,
%\end{equation}
%if assuming the host galaxy to be spherical (with an geometrical adjustment factor $\mathcal{G}$ to account for a range of morphologies), and for characteristic galaxy size $R$. $\mathcal{A}^{\rm ISM}(\epsilon)$ is introduced as the internal $\gamma$-ray attenuation, which is dominated by $\gamma\gamma$ interactions with reprocessed starlight by dust.
%
%Previous studies looking at modelling $\gamma$-ray emission from starburst galaxies:~\cite{Pfrommer2017ApJ}
%
%Other possibly relevant stuff:~\cite{Peng2019AA}

\subsection{Cosmological propagation and reprocessing of $\gamma$-rays}
\label{sec:cosmological_gamma_propagation}

To form the $\gamma$-ray background as we observe it from Earth at $z=0$, high energy photons emitted from source populations must propagate through intergalactic space over cosmological distances.
During their propagation, $\gamma$-ray photons interact with soft EBL photons at IR and optical wavelengths. Fundamentally, this is the same process as that which leads to the attenuative losses of $\gamma$-rays within a SFG
(considered in section~\ref{sec:internal_attenuation}). However, after their initial formation via pair-production~\citep{Heitler1954book, Madau1996ApJ},
they Compton up-scatter EBL and CMB photons
to form the diffuse secondary flux of $\gamma$-rays~\citep[e.g.][]{Wang2011ApJ, Inoue2013ApJ, Lacki2014ApJ} which continues to propagate and interact if photon energies remain sufficient. This \textit{cascade} reprocessing is coupled with the concurrent red-shifting of 
the $\gamma$-ray beam, which can be modelled using a covariant radiative transfer approach. We differentiate between dimensionless energies for soft EBL or CMB photons and $\gamma$-ray photons using the notation $\epsilon$ and $\epsilon_{\gamma}$ respectively. Moreover, $\gamma_{\rm e} = E_{\rm e}/m_{\rm e} c^2$ is introduced as the electron Lorentz factor.

\subsubsection{Cosmological $\gamma$-ray radiative transfer}

The propagation of $\gamma$-rays through soft intergalactic radiation fields may be modelled using a radiative transfer approach, where $\gamma$-ray emitting populations form the source function, while the cascade process effectively operates as an attenuation process at high energies, and an emission process at lower energies. 
Over cosmological distances, the radiative transfer equation, in terms of redshift, takes the form:
\begin{equation}
\frac{{\rm d}\mathcal{I}_{\gamma}}{{\rm d}z} = (1+z) \left[-\alpha_{\gamma \gamma} \mathcal{I}_{\gamma} + \frac{j_{\gamma}}{\nu^3} \right] \frac{{\rm d}s}{{\rm d}z}
\label{eq:radiative_transfer}
\end{equation}
\citep[e.g.][]{Chan2019MNRAS}\footnote{This follows from the covariant approach introduced by~\citealt{Fuerst2004AA}, which ensures conservation of photon number and phase space volume.}
%(see also~\citealt{Younsi2012AA}). This is adopted as it 
where all quantities are Lorentz invariant, i.e. $\mathcal{I}_{\gamma} = I_{\gamma}/v^3$ for $I_{\gamma}$ as the local `proper' intensity (such that, in practice, co-moving absorption $\alpha_{\gamma \gamma}$ and emission $j_{\gamma}$ functions are used for the attenuation and cascade re-emission of $\gamma$-rays respectively, as well as co-moving frequency $\nu$), and ${{\rm d}s}/{{\rm d}z}$ for a flat Friedmann-Robertson-Walker (FRW) Universe is given by
\begin{equation}
 \frac{{\rm d}s}{{\rm d}z} = \frac{c}{H_0\;\!(1+z)}\;\!\left( \Omega_{\rm r, 0}(1+z)^4 + \Omega_{\rm m, 0}(1+z)^3 + \Omega_{\rm \Lambda, 0} \right)^{-1/2}
\end{equation}
\citep[see, e.g.][]{Peacock1999_book}, where $\Omega_{\rm m,0} = 0.315\pm0.007$, $\Omega_{\rm r,0} \approx 0$ and $\Omega_{\rm \Lambda,0} = 0.685\pm0.007$ are the normalised density parameters for matter, radiation and dark energy respectively, and $H_0 = 100 h ~\text{km}\;\!\text{s}^{-1}\;\!\text{Mpc}^{-1}$ is the present value of the Hubble constant, where $h = 0.673\pm 0.006$~\citep{Planck2018}. We solve equation~\ref{eq:radiative_transfer} by discretising a source SFG population into redshift shells. The discretised solutions are then integrated over redshift, to find the total EGB intensity from $z=0$ to a maximum redshift, $z = z_{\rm max} = 3$, a range which covers the peak of cosmic star-formation~\citep{Madau2014ARAA} and would  presumably account for the majority of $\gamma$-ray emission from SFGs.

\subsubsection{$\gamma$-ray absorption and cascade reprocessing}

The absorption of $\gamma$-rays by cascade pair-production in the EBL can be characterised by an absorption coefficient,
%\begin{equation}
%\alpha(z) = \;\! \int_{\epsilon_{\rm th}}^{\infty} \frac{{\rm d}\epsilon}{\epsilon^{2}} \;\! n_{\rm ph}(\epsilon; z') \;\!\varphi\left(\epsilon \epsilon_{\gamma}[1+z']\right) \ ,
%\end{equation}
\begin{equation}
\alpha_{\gamma \gamma}(z', \epsilon_{\gamma}) = \frac{\sigma_{\rm T}}{\epsilon_{\gamma}^2}\;\! \int_{1/\epsilon_{\gamma}}^{\infty} {\rm d} \epsilon \;\! \epsilon^{-2} \;\! n_{\rm ph} (\epsilon; z')  \varphi(\epsilon^{\star}) \ ,
\label{eq:abs_coeff_gamma_gamma}
\end{equation}
(\citealt{Nikishov1961JETP}; see also~\citealt{Gould1967PhRv, Brown1973ApL}), 
where $\epsilon^{\star}$ retains its earlier definition of $\epsilon \epsilon_{\gamma}$, and
$\varphi(\epsilon^{\star})$ is given by equation~\ref{eq:varphi_def}. Over cosmological distances, a corresponding $\gamma$-ray optical depth due to pair-production may be written as
\begin{equation}
\tau_{\gamma \gamma}(z, \epsilon_{\gamma}) \equiv \;\!\int_0^z \alpha_{\gamma \gamma}(z', \epsilon_{\gamma}) \;\! \frac{{\rm d}s}{{\rm d}z'}{\rm d}z'  \ .
\end{equation}
The EBL and its impact on $\gamma$-ray absorption has been extensively studied (see~\citealt{Dwek2013APh, Cooray2016RSOS} for reviews) via direct measurements in UV/optical and/or near-IR bands~\citep[e.g.][]{Matsuoka2011ApJ, Berta2011AA, Bethermin2012AA, Driver2016ApJ, Andrews2018MNRAS}, indirect measurements using the attenuation of high-energy $\gamma$-rays from extra-galactic sources~\citep[e.g.][]{Desai2019ApJ, Abeysekara2019ApJ, Pueschel2019ICRC, Acciari2019MNRAS}, and theoretical models.

EBL models typically follow one of three approaches:
(1) forward-evolutionary models, which convolve spectral models with cosmic star-formation histories to estimate the EBL's development over redshift~\citep[e.g.][]{Kneiske2010AA, Finke2010ApJ}; (2) backward-evolution models, which extrapolate observed properties of galaxies in the local Universe to higher redshifts~\citep[e.g.][]{Franceschini2008AA, Dominguez2011MNRAS, Helgason2012ApJ, Stecker2012ApJ, Franceschini2017AA}, and (3) semi-analytical models, SAMs (see~\citealt{Kauffmann1993MNRAS, Cole1994MNRAS}) of hierarchical galaxy formation~\citep[e.g.][]{Gilmore2009MNRAS, Younger2011MNRAS, Gilmore2012MNRAS, Inoue2013ApJ}. Of these, forward-evolution models (category 1) suffer from several drawbacks. Notably, they do not trace the detailed evolution of crucial quantities which can impact the EBL spectrum, they are not able to reproduce certain observables (e.g. the observed rate of core-collapse supernovae -- see~\citealt{Horiuchi2011ApJ}), and that they may be based on over-estimated measures of stellar mass densities~\citep{Kobayashi2013ApJ}.
%\footnote{Typically, most models are based on the cosmic star-formation histories from~\citet{Hopkins2004ApJ} and~\citet{Hopkins2006ApJ}.}
Alternatively, while backward-evolutionary models (category 2) offer a robust EBL model at low and intermediate redshifts, they experience increased uncertainties at high redshifts. 

The final category of models, based on hierarchical galaxy formation SAMs, are the most detailed. They account for quantities such as halo merger histories, star-formation, feedback, gas cooling and chemical enrichment over large redshift ranges, providing properties of galaxies that are consistent with observations to relatively high redshifts~\citep[e.g.][]{Somerville2001MNRAS, Nagashima2004ApJ, Kobayashi2007ApJ, 
Kobayashi2010ApJ, Somerville2012MNRAS}. In this work, we adopt the SAM-based model of~\citet{Inoue2013ApJ}. It is based on the hierarchical galaxy formation model of~\citet{Nagashima2004ApJ}, which has been found to reproduce luminosity functions, luminosity densities and stellar mass densities of galaxies, as well as the luminosity functions of Lyman-break galaxies and Lyman-$\alpha$ emitting galaxies up to $z\sim6$~\citep{Kobayashi2007ApJ, Kobayashi2010ApJ}. 

The~\citet{Inoue2013ApJ} $\gamma$-ray optical depths for the EBL are provided between energies of 1 GeV and 45 TeV, up to a maximum redshift of $z=10$. This redshift range far exceeds our requirements (i.e. $z_{\rm max} = 3$), however, in a very small number of cases we use a logarithmic extrapolation to energies above and below the original range when necessary. At these energies the optical 
depth is low, so the impact of these points on our results is negligible.
We use this to compute the $\gamma$-ray absorption coefficient, $\alpha_{\gamma\gamma}$ (equation~\ref{eq:abs_coeff_gamma_gamma}) from the differential optical depth as a function of redshift. This is also used to calculate the secondary $\gamma$-ray emission, which relies on $\gamma$-ray absorption for the production of intermediate electrons. 
These electrons are injected along a $\gamma$-ray beam with a spectral number density of 
\begin{equation}
%(\gamma_{\rm e}, z, \epsilon_{\gamma})
\frac{{\rm d}n_{\rm e}}{{\rm d}\gamma_{\rm e}} \approx   \frac{2}{\epsilon_{\gamma} c}\int_{z}^{z_{\rm max}}\;\!{\alpha_{\gamma\gamma}(z', \epsilon_{\gamma})\;\!{I}_{\gamma}(z', \epsilon_{\gamma})} \;\! \frac{{\rm d}s}{{\rm d}z'} \;\! {\rm d}z' \ ,
\label{eq:electrons_rate}
\end{equation}
%NB minus sign from choosing alpha to be positive
where we approximate $\epsilon_{\gamma} \approx 2\gamma_{\rm e}$ (cf the delta function approximation of~\citealt{Boettcher1997AA}). This holds when the $\gamma$-ray energies are much greater than the energies of the soft EBL photons, and when $\gamma_{\rm e} \gtrsim 10^3$. Their resulting secondary $\gamma$-ray emission (from Compton up-scattering of primarily CMB photons, which dominate the energy density of the background radiation fields)
%\footnote{While EBL photons would also contribute, their contribution is relatively unimportant, given their energy density is much lower than the CMB by around two orders of magnitude.} 
can then be calculated by
\begin{align}
%(\gamma_{\rm e}, z, \epsilon_{\gamma})
j_{\gamma} &= \frac{{\rm d}}{{\rm d}t}\frac{{\rm d}n_{\gamma}}{{\rm d}\epsilon_{\gamma}} \nonumber \\
& = \frac{3 \sigma_{\rm T} c}{4} \int_{\gamma_{\rm e, min}}^{\gamma_{\rm e, max}} \frac{{\rm d}\gamma_{\rm e}}{\gamma_{\rm e}^2}\;\!\frac{{\rm d}n_{\rm e}}{{\rm d}\gamma_{\rm e}}\;\! \int_{0}^{1}\;\!{\rm d}x_c\;\!n_{\rm ph}(x_c, z) \;\! {f(x_c)}\;\!x_c^{-1} \ ,
\label{eq:emission_function}
\end{align}
% = \frac{3 \sigma_{\rm T} c}{4\gamma^2 \epsilon} \int_{\gamma_{\rm e}} {\rm d}\gamma_{\rm e}\;\!\frac{{\rm d}n_{\rm e}(z, \gamma_{\rm e})}{{\rm d}\gamma_{\rm e}}\;\! \int_{0}^{1}\;\!{\rm d}x\;\!\frac{{\rm d} n_{\rm ph}(x, z)}{{\rm d}x} \;\! f(x) \;\! x^{-1} \nonumber \\
%(nb. j is a differential quantity per $\epsilon_{\gamma}$ bin).
as required for equation~\ref{eq:radiative_transfer}, assuming that inverse Compton scattering takes place in the Thomson limit.
%\footnote{See also~\citet{Kneiske2008AA, Inoue2012PhRvD, DiMauro2014ApJ}, which adopt a slightly different approach.} 
Here, we use $f(x_c) = 2 x_c \ln x_c + x_c + 1 - 2 x_c^2$ (for $0<x_c<1$), and the dimensionless variable $x_c = \epsilon_{\gamma}/(4\gamma_{\rm e}^2 \epsilon)$
~\citep{Blumenthal1970RvMP, Rybicki1979_book}.
We set $\gamma_{\rm e, min} = 1$ and $\gamma_{\rm e, max} = \epsilon_{\rm \gamma, max}/2$, which we practically take as $\gamma_{\rm e, max} = \gamma_{\rm p}^{\star} m_{\rm p} c^2 /6$.

\section{Populations of star-forming galaxies}
\label{sec:section3}

In section~\ref{sec:cr_flux_normalisation}, it was shown that the $\gamma$-ray luminosity of a galaxy can be largely specified by its supernova (SN) event rate, $\mathcal{R}_{\rm SN}$. It was shown that this is directly related to the star-formation rate, $\mathcal{R}_{\rm SF}$ if assuming an IMF ($\mathcal{R}_{\rm SN} \approx 0.05 \;\!\mathcal{R}_{\rm SF}$ for a Salpeter IMF,~\citealt{Salpeter1955ApJ}) so, if a population of SFGs can be characterised by the distribution of its star-formation rates, its redshift distribution, and its spatial clustering characteristics, the $\gamma$-ray luminosity and spatial emission properties of that population can be modelled.

\subsection{Star-formation rates}
\label{sec:sfrf}

The star-formation rate function, $\Psi(\mathcal{R}_{\rm SF}, z)$ (SFRF)
is the number density of galaxies as a function of their star-formation rate. 
Its evolution is determined by
the underlying history of galaxy assembly, together with 
gas cooling, feedback (from AGN and stars/stellar end-products) and prior star-formation
within galaxies.
As such,
modelling the SFRF reliably has proven to be
a challenging task.
To date, 
various approaches have been adopted, including SAMs~\citep{Fontanot2012MNRAS, Gruppioni2015MNRAS}
and hydrodynamic simulations~\citep{Dave2011MNRAS, Tescari2014MNRAS, Katsianis2017MNRAS_b}, with varying degrees of success. Indeed,
many previous approaches have been found to  
yield higher numbers of galaxies at all SFRs compared to observations 
 -- a discrepancy often attributed to limitations in the implementation 
of feedback physics~\citep[see][for further discussion]{Tescari2014MNRAS, Katsianis2017MNRAS_b}.

In this work, we adopt a SFRF reference model of~\citealt{Katsianis2017MNRAS}, which is obtained from simulations using Virgo Consortium's Evolution and Assembly of GaLaxies and their Environments (EAGLE) project~\citep{Schaye2015MNRAS, Crain2015MNRAS}. We used their reference model  (\verb|100N1504-Ref|)
 as it offered coverage of a large range of SFRs (around $10^{-3}$ to $10^3~\text{M}_{\odot}~\text{yr}^{-1}$) up to redshifts up to $z\sim8$.
When compared with observationally-determined SFRFs , as discussed in~\citealt{Katsianis2017MNRAS}, this was found to under-predict the number of galaxies with SFRs of 1-10 ${\rm M}_{\odot}\;\!{\rm yr}^{-1}$at $z>3$, and the number of objects with SFRs of 10-100 ${\rm M}_{\odot}\;\!{\rm yr}^{-1}$ at $z<2$.\footnote{Comparison of the~\citealt{Katsianis2017MNRAS} reference model is made with SFRFs constructed from UV, IR H$\alpha$ and radio luminosity functions to facilitate broad SFR and redshift coverage. Observationally-derived SFRFs from~\citet{Mauch2007MNRAS, Reddy2008ApJS, Gilbank2010MNRAS, Rodighiero2010AA, Karim2011ApJ, Ly2011ApJ, Robotham2011MNRAS, Gruppioni2013MNRAS, Magnelli2013AA, Patel2013MNRAS, Sobral2013MNRAS, Bouwens2015ApJ, Alavi2016ApJ, Marchetti2016MNRAS, Parsa2016MNRAS} as well as from compiled data~\citep{Madau2014ARAA}, are used. Additionally, comparison is made with SFRFs from~\citet{Smit2012ApJ, Duncan2014MNRAS, Katsianis2017MNRAS_b}.} We note that the~\citealt{Katsianis2017MNRAS} model assumes 
a~\citet{Chabrier2003PASP} IMF, but a~\citet{Salpeter1955ApJ} IMF is adopted in our calculation for the $\gamma$-ray luminosity of a galaxy (equation~\ref{eq:cr_luminosity}) and the luminosity of its dust emission (equation~\ref{eq:dust_conversion}).
If a Salpeter IMF had been assumed, the resulting SFRs would roughly be a factor of 1.8 higher~\citep{Katsianis2017MNRAS}. 
To correct for this discrepancy, we therefore scale the SFRF model accordingly.

Integrating over the SFRF yields the cosmic star-formation rate density (CSFRD),
\begin{equation}
\rho^{\star}(z) = \int \Psi(\mathcal{R}_{\rm SF}, z) \;\! {\rm d} \log_{10} \mathcal{R}_{\rm SF}
\end{equation}
where $\Psi(\mathcal{R}_{\rm SF}, z)$ is the SFRF in units of ${\rm Mpc}^{-3}$ per decade in $\mathcal{R}_{\rm SF}$. \citet{Katsianis2017MNRAS} demonstrated that 
a CSFRD function derived from the baseline \verb|100N1504-Ref| model 
was 
found the exhibit a consistently lower normalisation
than that from observation by a factor of 1.5, which may result from the differences with respect to observations discussed above. To account for this, we apply a further multiplicative correction to our model.
The CSFRD function derived from the \verb|100N1504-Ref| model was
otherwise largely consistent with observations~\citep{Gilbank2010MNRAS, Karim2011ApJ, Robotham2011MNRAS, Sobral2013MNRAS, Madau2014ARAA, Bouwens2015ApJ}
with the exception of that obtained from IR data~\citep{Rodighiero2010AA, Madau2014ARAA}. This was considered to be due to assumed dust corrections in computing UV luminosities, incomplete UV luminosity functions or possible overestimations of the SFR from IR data~\citep{Katsianis2017MNRAS}.
It was further shown that those SFGs with high star-formation rates, between 10 and 100 $\text{M}_{\odot}\;\!\text{yr}^{-1}$, exhibit the strongest redshift dependence, peaking sharply at $z\sim 2$, while less vibrantly star-forming galaxies show a weaker evolution in their contribution to the CSFRD (this is also in tension with IR studies, e.g. ~\citealt{Magnelli2013AA}, which do not find such a strongly peaked evolution of highly star-forming galaxies).
%\footnote{It was noted that this is also in tension with IR studies~\citep{Magnelli2013AA}, which do not find such a strongly peaked evolution of highly star-forming galaxies.}
These intensively star-forming galaxies presumably represent the most important SFG sub-class contribution to the EGB, which should also reflect this strongly peaked evolutionary history. It will be shown in the following sections (\ref{sec:power_spec} and~\ref{sec:anisotropies}) that this would imprint a distinctive spatial signature into the EGB.

\subsection{Clustering and bias}
\label{sec:power_spec}

In the hierarchical model of structure formation, spatial clustering of galaxies is primarily determined by the distribution of dark matter in the Universe. 
Dark matter haloes form from the gravitational collapse of primordial Gaussian density perturbations, with their development and properties having been well-studied through $N$-body simulations and analytic models~\citep[e.g.][]{Springel2005Natur, Reed2009MNRAS, Jose2016MNRAS, Jose2017MNRAS}. 
The clustering properties of dark matter haloes are strongly influenced by the matter power spectrum of the Universe and, by extension, the cosmological parameters~\citep[e.g.][]{Hu1998ApJ, Eisenstein1999ApJ, Jose2013MNRAS}.
Galaxies emerge in virialised dark matter haloes through gas cooling~\citep{Rees1977MNRAS}, with their formation efficiency being governed by their virial temperature and gas density (which are influenced by the gravitational potential, and hence mass, of the halo -- see~\citealt{Silk1993PhR, Sutherland1993ApJS}). The subsequent evolution of galaxies through cosmic time experiencing accretion of new gas from the cosmic web, feedback and mergers yields the properties of populations of galaxies at high redshifts~\citep{Jose2013MNRAS, Harikane2018PASJ} and, eventually in the present Universe~\citep[e.g.][]{Press1974ApJ, Lacey1993MNRAS, Sheth1999MNRAS, Behroozi2013ApJ} and so form biased tracers of the underlying dark matter distribution of the Universe at different epochs~\citep[e.g.][]{Kaiser1984ApJ, Cooray2002PhR, Mo2010book}. 

The bias of galaxy population clustering compared to that of dark matter is typically studied observationally from their spatial distribution, with various sources classes having been found to exhibit different clustering properties~\citep[e.g. see][which finds a different clustering bias for AGNs and SFGs against dark matter, with AGNs typically exhibiting greater clustering strength]{Hale2018MNRAS}. We define the effective clustering bias factor of SFGs compared to dark matter using the relation $P_{\rm g}(k, z) = b_{\rm SFG}(z)\;\!P_{\rm lin}(k, z)$, where 
$P_{\rm g}(k, z)$ is the power spectrum of SFGs,
and $P_{\rm lin}(k, z)$ is the power spectrum of linear dark matter density fluctuations. We calculate $P_{\rm lin}(k, z)$ using the transfer function 
%\footnote{The transfer function is defined as the ratio of the time-integrated growth on a length-scale (or comoving wavenumber, $k$) compared to growth on scales far larger than the Jeans scale.} 
approximation of~\citet{Eisenstein1999ApJ}, which is shown to be accurate to within 5\%. 

The SFG population bias factor, $b_{\rm SFG}$ may be calculated from the ratio of galaxy to dark matter correlation functions, i.e:
\begin{align}
b_{\rm SFG}^2(z) & = \frac{\xi_{\rm g}(r, z)}{\xi_{\rm DM}(r, z)} \nonumber \\
& = \left( \frac{r_0(z)}{8} \right)^{\iota} \;\! \frac{J_2}{\sigma_8^2 \; \mathcal{G}^2(z)}
\label{eq:bias_func}
\end{align} 
\citep[e.g.][]{Kaiser1984ApJ, Bardeen1986ApJ, Lindsay2014MNRAS}, where the matter fluctuation amplitude $\sigma_8 = 0.811\pm0.006$~\citep{Planck2018}, 
and $\mathcal{G}(z) = g(z)/g_0$, with
 $g(z)$ as the growth factor at redshift $z$ and $g_0 = g(z=0)$~\citep[e.g.][]{Carroll1992ARAA}.\footnote{We calculate this using the formula presented in~\citealt{Hamilton2001MNRAS}, using the public code provided at:~\url{https://jila.colorado.edu/~ajsh/growl/}.} Additionally, $J_2 = 72/([3-\iota][4-\iota][6-\iota]2^{\iota})$, and 
$r_0(z) = r^{\rm c}_0 (1+z)^{p}$ with $p = 1-({3+w})/{\iota}$~\citep{Lindsay2014MNRAS}.
Here, the choice of the parameter $w$ reflects the clustering model adopted. In this demonstrative model we consider only \textit{linear clustering}~\citep{Overzier2003AA} where clustering growth is set by linear perturbation theory and $w = \iota-1$. We leave the investigation of alternative clustering growth models to future work -- for example, \textit{stable clustering} (where clusters have a fixed physical size and $w = 0$), \textit{co-moving clustering} (where clusters have fixed co-moving size and $w = \iota -3$) and \textit{decaying clustering} (which implies a rapid clustering decay) are also considered in the literature~\citep{Overzier2003AA, Kim2011MNRAS, Elyiv2012AA}. 
The remaining parameters in equation~\ref{eq:bias_func} are the power-law slope of the two-point correlation function of galaxies, $\iota$, and the galaxy clustering length $r_0^{\rm c}$. Both of these may be estimated empirically for SFGs, and we adopt the best-fit values of~\citet{Hale2018MNRAS}: $\iota = 1.8$ and
$r_0^{\rm c} = 6.1\;\!\text{Mpc}\;\!h^{-1}$. These were computed from
radio-selected SFGs in the COSMOS field using deep Karl G. Jansky Very Large Array (VLA) data at 3 GHz, reaching redshifts as high as $z\sim 5$, thus covering our range of interest ($z\leq 3$).\footnote{We note that~\citealt{Magliocchetti2017MNRAS} also provided values for these parameters for a radio-selected sample of SFGs at 1.4 GHz (with radio fluxes above 0.15 mJy) up to $z\sim3$. However, the number of data points in their analysis is much fewer than in~\citet{Hale2018MNRAS}, leading to our preference to use the best-fit values of the later study.}
The resulting bias factor from these parameter choices is higher than those computed for SFGs at other wavelengths~\citep[e.g.][]{Gilli2007AA, Starikova2012ApJ, Magliocchetti2013MNRAS}, but this is attributed to the greater extent of the redshift distribution of the sources.

\subsection{Development of EGB anisotropies}
\label{sec:anisotropies}

\begin{figure}
    \centering
    \includegraphics[width=0.9\columnwidth]{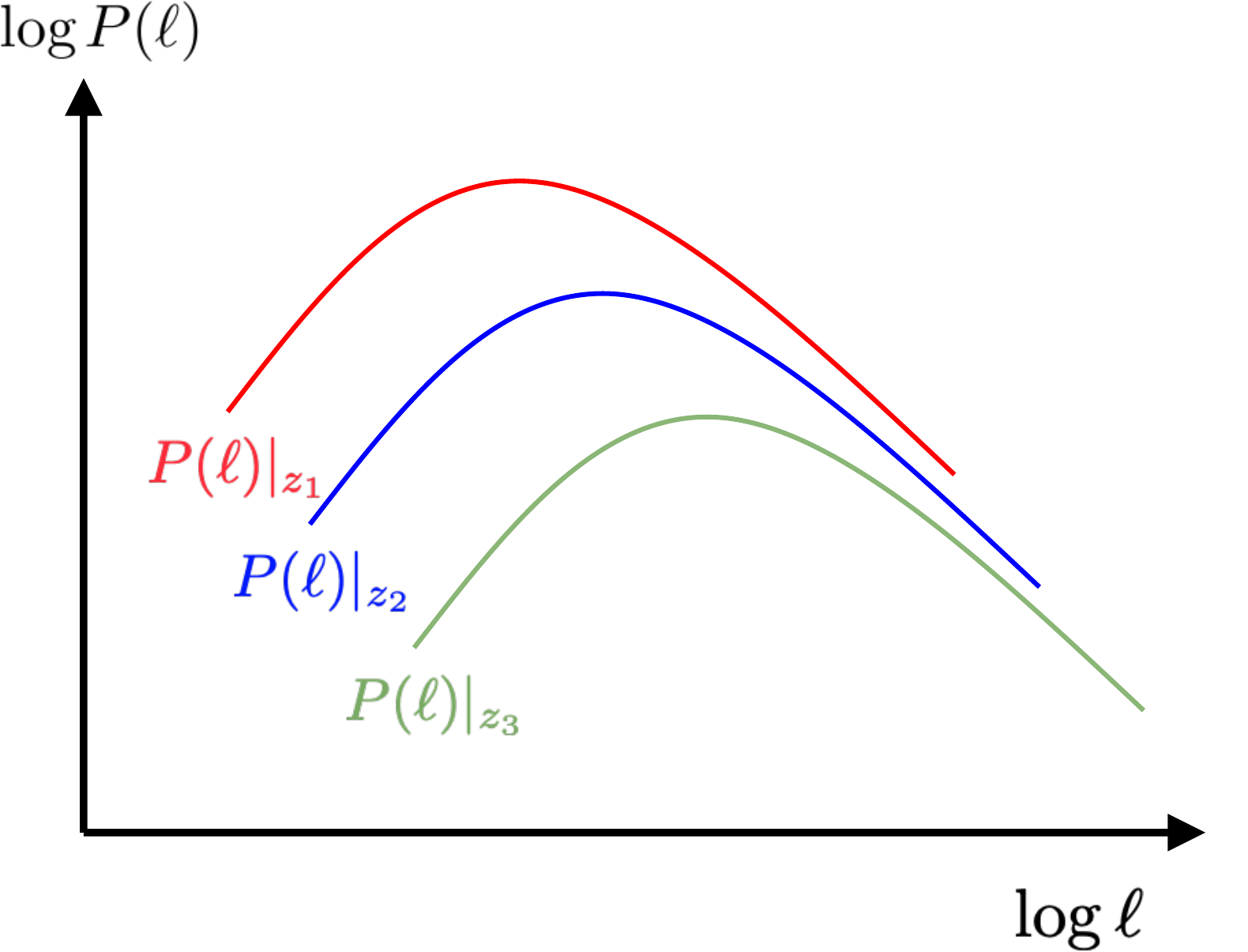}
    \caption{Schematic to illustrate the redshift evolution of the power spectrum of the source population as a function of multipole $\ell$. The peak would correspond to the multipole the dominant signal is imprinted at. Here, $z_1<z_2<z_3$.}
    \label{fig:pz_schematic}
\end{figure}
%The matter power spectrum of the Universe is usually defined as $P_{\rm m}(k) = \langle \tilde{\delta}_{\rm m}({\bf k})\;\!\tilde{\delta}_{\rm m}^{\star}({\bf k}')\rangle$, where $\tilde{\delta}_{\rm m}({\bf k})$ denotes the Fourier Transform of the matter over-density, $\delta({\bf x}) = \rho({\bf x})/\bar{\rho} - 1$, and ${\bf k}$ is the a comoving wave-vector of length $k$ (the comoving wavenumber corresponding to some co-moving spatial scale of $\ell = 2\pi/k$)

The SFG power spectrum $P_{\rm g}(k, z)$ would imprint a signature in the EGB, even though individual contributing sources would not typically be resolved. The distribution of spatial scales of this signature would depend on redshift $z$, being specified by $P_{\rm g}(k, z)$, and the strength of the contribution from a shell in redshift would corresponding to the relative $\gamma$-ray luminosity of the source population at that epoch, as set by the SFG redshift distribution (see schematic in Figure~\ref{fig:pz_schematic}). 
This could be measured from $\gamma$-ray background observations using the auto-correlation function (equation~\ref{eq:autocorr_term_app}), from which a
clustering term $\mathcal{C}_{\ell}^{C}$ and an 
isotropic Poisson noise term (an auto-correlation term) $\mathcal{C}_{\ell}^{P}$
 can be decomposed from the Fourier Transform (see Appendix~\ref{sec:clustering_power_spec} for details). These may be written as
\begin{equation}
\mathcal{C}_{\ell}^{C}(E_{\gamma}) =  \int_0^{z_{\rm max}} \frac{{\rm d}^2V_{\rm c}}{{\rm d}z\;\!{\rm d}\Omega}{\rm d}z\;\ P\left(\frac{\ell_p}{r_p}[1+z]\right)\left\{ \frac{{\rm d}F_{\gamma}(E_{\gamma}, z)}{{\rm d}E_{\gamma}} \right\}^2 \ ,
%\frac{{\rm d}\mathcal{C}_{\ell}^{\gamma}(E_{\gamma})}{{\rm d}E_{\gamma}} =  \int_0^{z_{\rm max}} \frac{{\rm d}^2V_{\rm c}}{{\rm d}z\;\!{\rm d}\Omega}{\rm d}z\;\ P\left(\frac{\ell}{r_p}\right)\left\{ \frac{{\rm d}F_{\gamma}(E_{\gamma}, z)}{{\rm d}E_{\gamma}} \right\}^2
\label{eq:differential_anisotropy}
\end{equation}
and
\begin{equation}
\mathcal{C}_{\ell}^{P}(E_{\gamma}) =  \int_0^{z_{\rm max}} \frac{{\rm d}^2V_{\rm c}}{{\rm d}z\;\!{\rm d}\Omega}{\rm d}z\;\! \left\{ \frac{{\rm d}F_{\gamma}(E_{\gamma}, z)}{{\rm d}E_{\gamma}} \right\}^2 \ ,
\label{eq:final_poisson}
\end{equation}
respectively, in differential units of flux, where 
$D_{\rm L}$ is the luminosity distance (equation~\ref{eq:lum_dist_app}) and
the flux term, ${\rm d}F_{\gamma}/{{\rm d}E_{\gamma}}$, accounts for the redshift-dependent emission of $\gamma$-rays from the population of SFGs, thus absorbing the internal and external $\gamma$-ray attenuation/reprocessing models, and the co-moving number density of SFGs.
 Our later results sum the contribution from equations~\ref{eq:differential_anisotropy} and~\ref{eq:final_poisson} to give the total angular power spectrum of the EGB from SFGs.

\section{Results and discussion}
\label{sec:section4}

\subsection{EGB spectrum}
\label{sec:intensity}

\begin{figure*}
    \centering
    \includegraphics[width=0.85\textwidth]{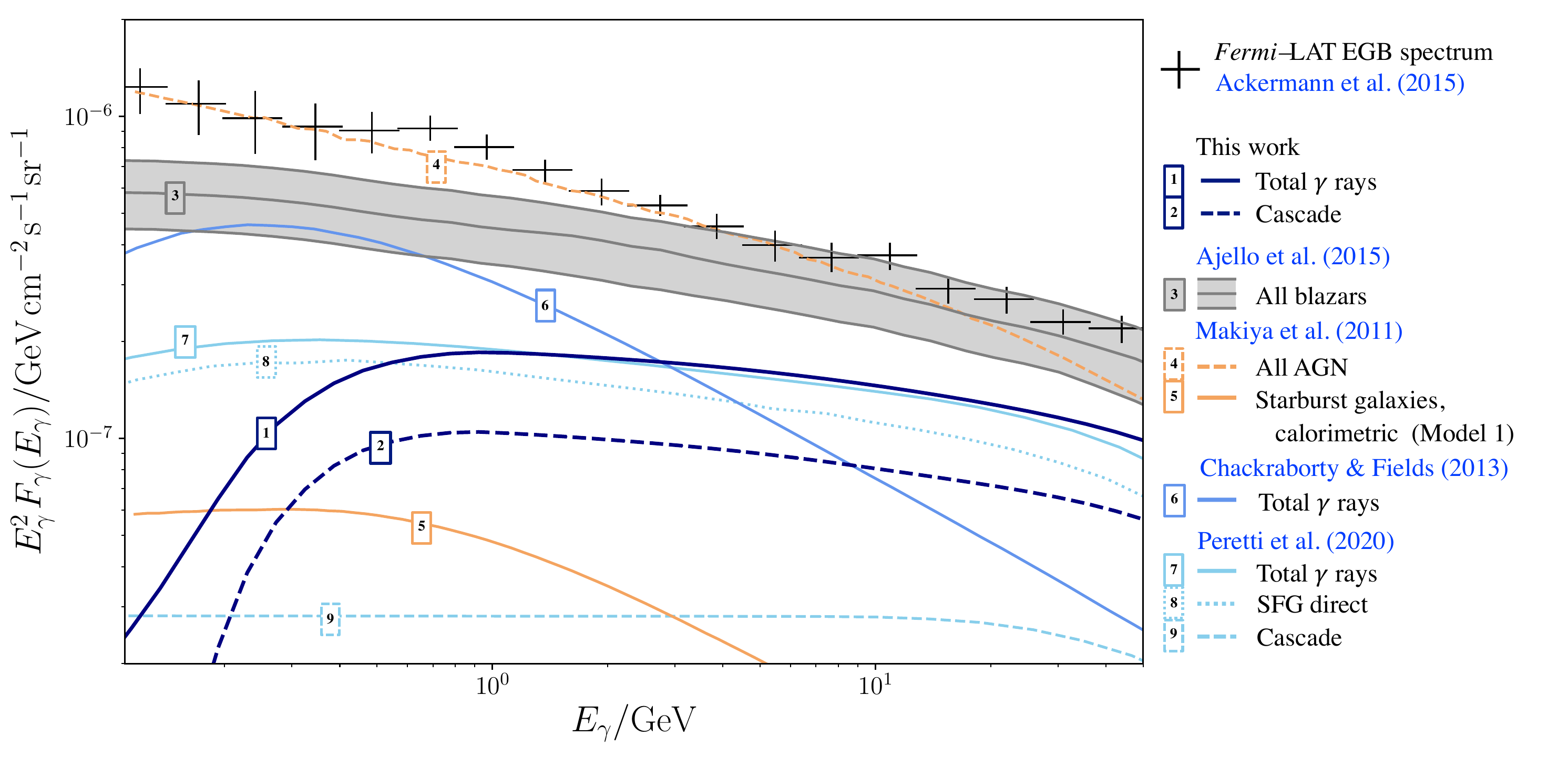}
    \caption{Total SFG contribution to the EGB at $z=0$ between $E_{\rm \gamma} = 0.1 - 50$ GeV (line 1), of which the contribution from cascade emission due to inverse-Compton scattered EBL radiation fields is indicated (line 2). Comparison with 50 months of \textit{Fermi}-LAT data for the the total
observed diffuse EGB (taken from~\citealt{Ajello2015ApJ}, with original data  from~\citealt{Ackermann2015ApJ}) is shown, together with model predictions for all blazar contributions (band 3, with the lower, middle and upper grey lines showing the three models presented in~\citealt{Ajello2015ApJ}), the all AGN and starburst galaxy contributions computed by~\citealt{Makiya2011ApJ} (lines 4 and 5, respectively), the total SFG contribution calculated by~\citealt{Chakraborty2013ApJ} (line 6), and the total, direct and cascade emission from SFG nuclei computed by~\citealt{Peretti2020MNRAS} (lines 7, 8 and 9, respectively).}
  \label{fig:EGB_spectrum}
\end{figure*}
The EGB spectrum between 0.1 and 50 GeV predicted at $z=0$ by our fiducial model, which adopts a characteristic SFG nucleus of $R = 0.1~{\rm kpc}$, a CR spectral index $\Gamma = 2.1$ and a maximum CR energy $E_{\rm max} = 50$ PeV, is shown in Figure~\ref{fig:EGB_spectrum}. Here, both the total contribution to the diffuse EGB from SFGs (line 1), and that arising from the cascaded SFG emission (line 2) are included. For comparison, the contribution from resolved and unresolved blazars is shown (band 3, denoting the range of 3 models presented in~\citealt{Ajello2015ApJ} -- however, these do not include a cascade flux component), together with the total
observed diffuse 
EGB spectrum using 50 months of \textit{Fermi}-LAT data (taken from~\citealt{Ajello2015ApJ}, with original data from~\citealt{Ackermann2015ApJ}). For reference, the contribution from all AGN presented in~\cite{Makiya2011ApJ} is also shown (line 4). \cite{Makiya2011ApJ} also compute the contribution from SFGs (line 5), which we find to be substantially lower than many other literature models.
It can be seen that our fiducial model is in agreement with the observational constraints given by the contribution to the EGB from resolved and unresolved blazars, 
however the predicted SFG contribution comes close to saturating the diffuse EGB at higher energies, above a few 10s GeV (but remains compatible with observational limits). This behaviour is also evident in some other models,  e.g.~\citealt{Peretti2020MNRAS} (line 7).

In Figure~\ref{fig:EGB_spectrum}, the substantial variation in predictions made by other models is clear. Here, we draw comparison between our fiducial model and those in the literature which consider a contribution specifically from SFGs.
We find our approach yields a $z=0$ EGB intensity that is much higher
than the 
 the SAM-based method considered by~\citealt{Makiya2011ApJ} (also that of \citealt{Lamastra2017A&A}, which falls substantially lower even than the \citealt{Makiya2011ApJ} prediction, and is not shown in  Figure~\ref{fig:EGB_spectrum}), which is exceeded by as much as an order of magnitude at energies above $\sim$ 10 GeV. Both the \citealt{Makiya2011ApJ} and \citealt{Lamastra2017A&A} SAM-based models are strongly dependent on the source population properties, redshift distributions and $\gamma$-ray emission models adopted, all of which differ compared to equivalent model components adopted in this work.
 
 By contrast, the SFG contribution intensity computed by~\cite{Chakraborty2013ApJ}, line 6, is substantially higher than our prediction. It also exceeds predictions by other models up to energies of $\sim$ 3 GeV, as shown. It is even comparable to the all blazar contribution of~\cite{Ajello2015ApJ} below $\sim$ 0.6 GeV. However, its steeper power-law in energy, resulting from the steeper assumed CR proton spectrum within the source population, causes the~\cite{Chakraborty2013ApJ} model to have fallen far below the prediction of this work by 50 GeV.
 
 The approach of~\cite{Peretti2020MNRAS}, line 7, is broadly consistent with the prediction of this work, with some deviations at lower energies and a smaller cascade contribution (line 9). 
 The low-energy difference is likely accounted for by the additional physics included in the spectral model of~\cite{Peretti2020MNRAS} that would boost the low-energy $\gamma$-ray flux compared to this work (for example, their inclusion of inverse-Compton and bremsstrahlung emission may become relatively important in lower star-formation rate sources, where pion-decay $\gamma$-ray emission would be less dominant). The differences in the cascade prediction between this work and that of~\cite{Peretti2020MNRAS} would presumably arise from their delta-function approximation of the EBL radiation field, compared our use of the~\citealt{Inoue2013ApJ} EBL model. Given the current uncertainties in EBL models, it can be reasonably argued that both approaches to the cascade emission are equally valid, and that future estimations of the cascade contribution will improve as observational constraints on the EBL are tightened.

\subsection{EGB anisotropy signatures}

% \begin{table}
% \centering
% \begin{tabular}{*{2}{c}}
% \hline \hline
% Parameter & Value \\
% \hline
% $R$ & $100\ \rm pc$ \\
% $\Gamma$ & 2.1 \\
% \hline \hline
% \end{tabular}
% \caption{A list of reference parameters 
%  used in the EGB fiducial characteristic galaxy model, and its internal hadronic CR spectrum.
%  These parameter choices are considered appropriate for a population of SFGs resembling those seen in the nearby Universe, with other unspecified quantities varying according to redshift and star-formation rate as determined by the treatment presented in~\cite{Katsianis2017MNRAS}.}
% \label{tab:param}
% \end{table} 

\begin{figure}
    \centering
    \includegraphics[width=\columnwidth]{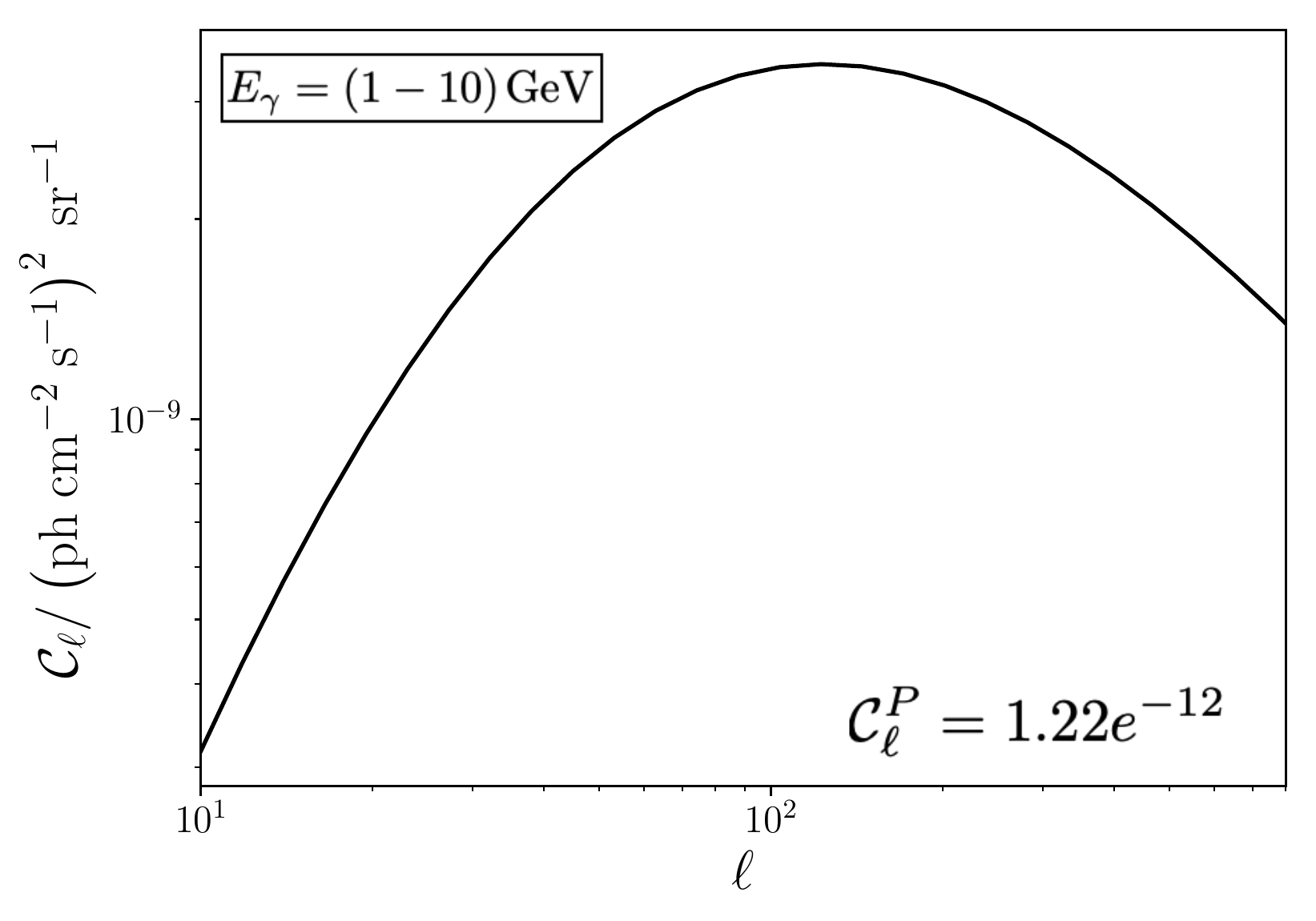}
    \caption{EGB intensity fluctuation angular power spectrum $\mathcal{C}_{\ell}$ shown against multipole $\ell$ in the energy band $E_{\gamma} = (1-10)~{\rm GeV}$, for the 
    fiducial model. $\mathcal{C}^{P}_{\ell}$ is the isotropic auto-correlation (Poisson noise) contribution, which is comparatively negligible. This adopts a characteristic SFG nucleus radius of $R = 0.1~{\rm kpc}$, an intrinsic source hadronic CR power-law spectrum of index $\Gamma = 2.1$, and a maximum CR energy of $E_{\rm max} = 50\;{\rm PeV}$.}
  \label{fig:baseline_anisotropies}
\end{figure}

We directly compute the EGB intensity fluctuation angular power spectrum at $z=0$ arising from our model SFG population. This uses the computational method outlined in  Appendix~\ref{sec:computational_method} to solve equations~\ref{eq:differential_anisotropy} and~\ref{eq:final_poisson}. 
Large numbers of photons are needed to compute high-resolution spectral statistics from data. Typically, $\gamma$-ray data analysis methods would bin events according to photon energy, to improve signal-to-noise ratios within an energy band and to reduce the requirement on photon numbers in a small energy range. We therefore compute our expected anisotropy signatures in broad energy bins to reflect this. Figure~\ref{fig:baseline_anisotropies} shows the EGB anisotropy signature computed for our fiducial model, integrated over the energy band $E_{\gamma} = (1-10)$ GeV.
%, thus covering the peak of the EGB flux (see Figure~\ref{fig:EGB_spectrum}).}
%Relevant parameter choices for the fiducial model are summarised in Table~\ref{tab:param}.
Uncertainties from the empirical dust relation of equation~\ref{eq:dust_relation} were propagated, but found to be negligible. While the total EGB anisotropy signature is plotted in this case, the clustering contribution (cf. equation~\ref{eq:sum_contributions}) exceeds the Poisson component by around 3 orders of magnitude -- consistent with the expectation that the Poisson (statistical noise) contribution from a source population comprised of a large number of unresolved faint galaxies would be relatively low.

\subsubsection{Energy bands}

\begin{figure}
    \centering
    \includegraphics[width=\columnwidth]{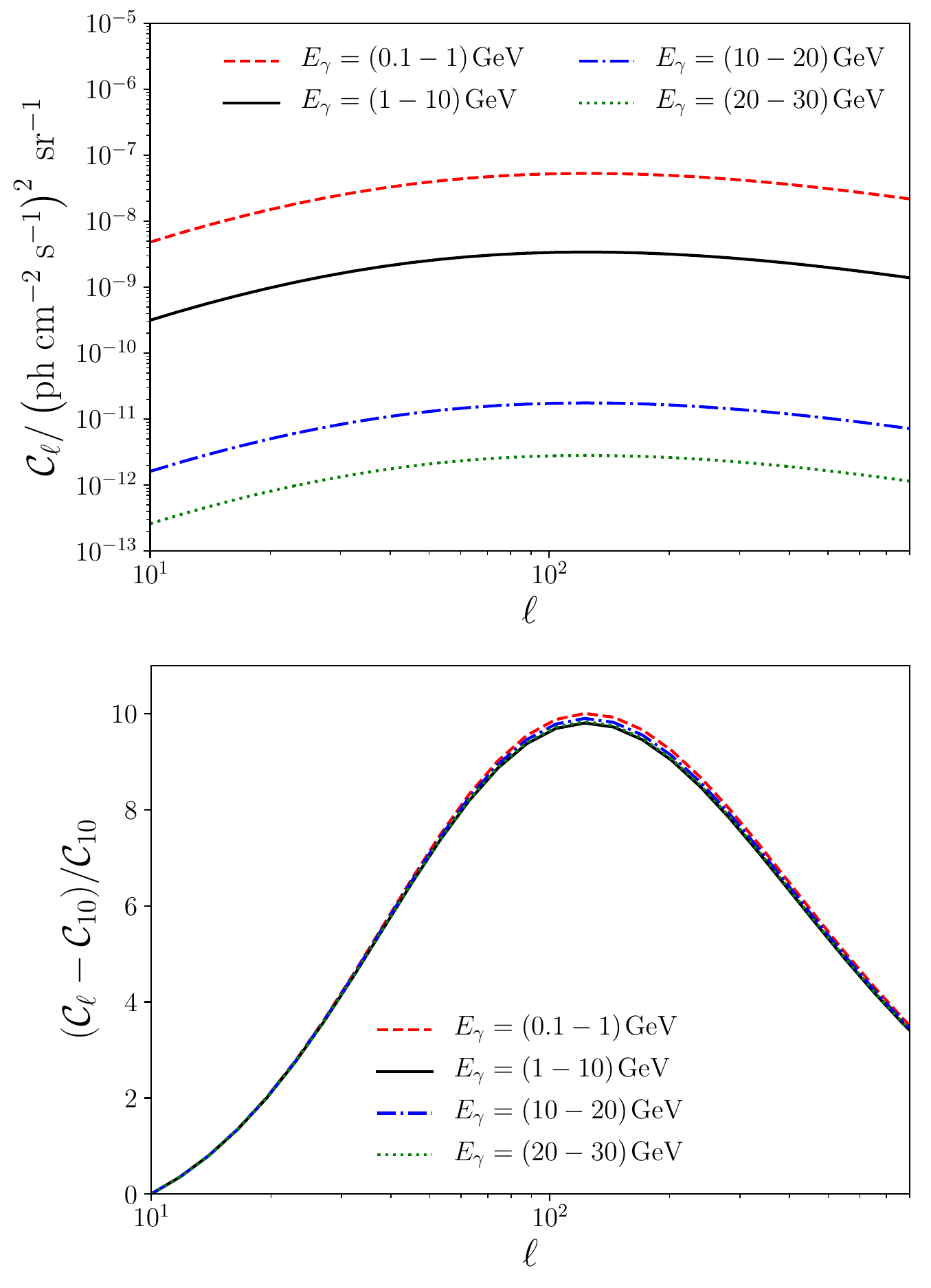}
    \caption{EGB intensity fluctuation angular power spectrum $\mathcal{C}_{\ell}$ shown against multipole $\ell$ in energy bands (top), with a normalised comparison in the bottom panel, to allow the shapes of the angular power spectra in the four bands to be compared.}
  \label{fig:comparison_energy}
\end{figure}

The intensity of the EGB varies with energy (cf. Figure~\ref{fig:EGB_spectrum}). 
As the cosmological attenuation of $\gamma$-rays is also energy-dependent, with stronger flux suppression arising at higher energies~\citep[e.g.][]{Gilmore2009MNRAS, Inoue2013ApJ}, the EGB anisotropy would differ according to the choice of energy band.
Figure~\ref{fig:comparison_energy} demonstrates that such differences are almost negligible, when comparing the EGB angular power spectrum in four bands, (0.1-1.0) GeV, (1.0-10) GeV, (10-20) GeV and (20-30) GeV for the fiducial model. The upper panel shows the main difference between these four energy bands follows simply from the EGB energy spectrum (Figure~\ref{fig:comparison_energy}). 
To remove this spectral energy dependence, the lower panel 
renormalises the $\mathcal{C}_{\ell}$s 
 relative to an arbitrary reference (taken here as $\mathcal{C}_{10}$). This allows the \textit{shape} of the anisotropy power spectrum in the four energy bands to be compared.
From this, some minor differences emerge, with a slightly broader spectral peak in the (1-10) GeV energy band compared to the others. Residuals between the (0.1-1) GeV, (10-20) GeV and (20-30) GeV bands compared to the (1-10) GeV band reveal a slight boost at larger scales and around the spectral peak for the (0.1-1) GeV band. This can be attributed to the cascade process: the (0.1-1) GeV band is not strongly affected by $\gamma$-ray attenuation, however it does receive a proportionally greater fraction of its photons than the other bands from cascaded $\gamma$-rays, which originate from higher energies and more distant sources. The cascaded contribution from these more distant sources is manifested as additional flux on larger scales. 
Conversely, the upper two energy bands would suffer more severely from $\gamma$-ray flux attenuation, and this would be more important compared to cascaded photons reprocessed into these bands. This would disproportionately affect $\gamma$-rays imprinted by more distant sources at larger angular scales, slightly reducing power at small-$\ell$s compared to lower energy bands, and causing the observed sharpening and slight skew in the EGB power spectrum for these bands.\\

\subsubsection{Model parameters}
\label{sec:model_params}

\begin{figure}
    \centering
    \includegraphics[width=\columnwidth]{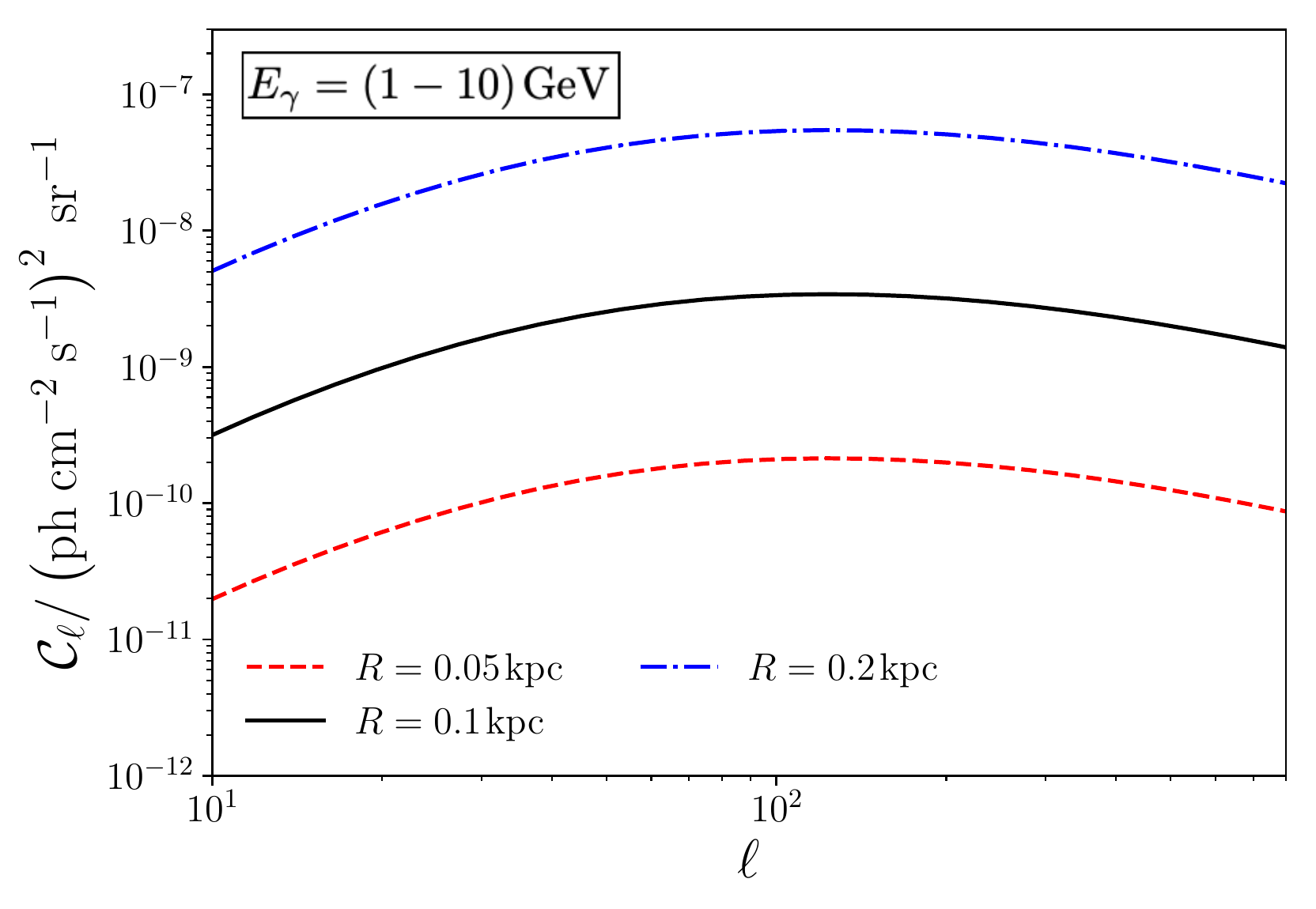}
    \caption{EGB intensity fluctuation angular power spectrum $\mathcal{C}_{\ell}$ shown against multipole $\ell$, for alternative choices of the characteristic starburst nucleus size of the source population from $R = 0.1~{\rm kpc}$ (fiducial value) to $R = 0.2~{\rm kpc}$ and 0.05 kpc. While there is a strong impact on the EGB intensity for difference choices of $R$, the angular power spectrum shape is not affected.  Results are shown for the energy band $E_{\gamma} = (1-10) \;{\rm GeV}$.}
  \label{fig:comparison_radius}
\end{figure}

The three fixed parameters in the fiducial model are $R$, $\Gamma$ and $E_{\rm max}$. However, some variation of their values would be expected throughout a real SFG source population, with implications for the EGB intensity and anisotropy. 
The radius $R$ of a SFG nuclear region could vary substantially between galaxies. For example, among starburst galaxies in the local Universe, it is found to differ by a factor of a few -- in NGC 253, $R\approx 0.1~{\rm kpc}$~\citep{Weaver2002ApJ}, while for M82, $R \approx 0.25~{\rm kpc}$~\citep{deGrijs2001}. Moreover, in models and simulation work, compact galaxies are found to be common at high-redshift~\citep[e.g.][]{Furlong2017MNRAS}, which would imply a redshift-dependence in $R$ for realistic SFG source distribution models. Such variations would have 
discernible effects on the EGB intensity and anisotropy. The impact of alternative choices of $R$, with the value increased and decreased by a factor of 2 compared to the fiducial choice of 0.1 kpc are shown in Figure~\ref{fig:comparison_radius}. This demonstrates the EGB intensity is directly affected by the value of $R$ set in the source population, with higher intensities developing for a larger characteristic choice of $R$. 
This effect can be understood from the spatial spread of photons through a SFG nucleus when $\mathcal{R}_{\rm SF}$ is fixed. Increasing $R$ would increase the volume of the SFG nucleus, and decrease the photon density in the stellar and dust radiation fields that attenuate $\gamma$-rays. More $\gamma$-rays would then escape from their source galaxy, contributing more photons to the EGB. Anisotropies are unaffected in this case, as $R$ is adjusted independently of redshift. If a more physical redshift-dependent treatment of $R$ were adopted, an anisotropic signature would presumably emerge in the EGB. However, the necessary detailed modelling of appropriate redshift-size relations for SFG populations falls beyond the scope of this study, and is left to future dedicated work.

\begin{figure}
    \centering
    \includegraphics[width=\columnwidth]{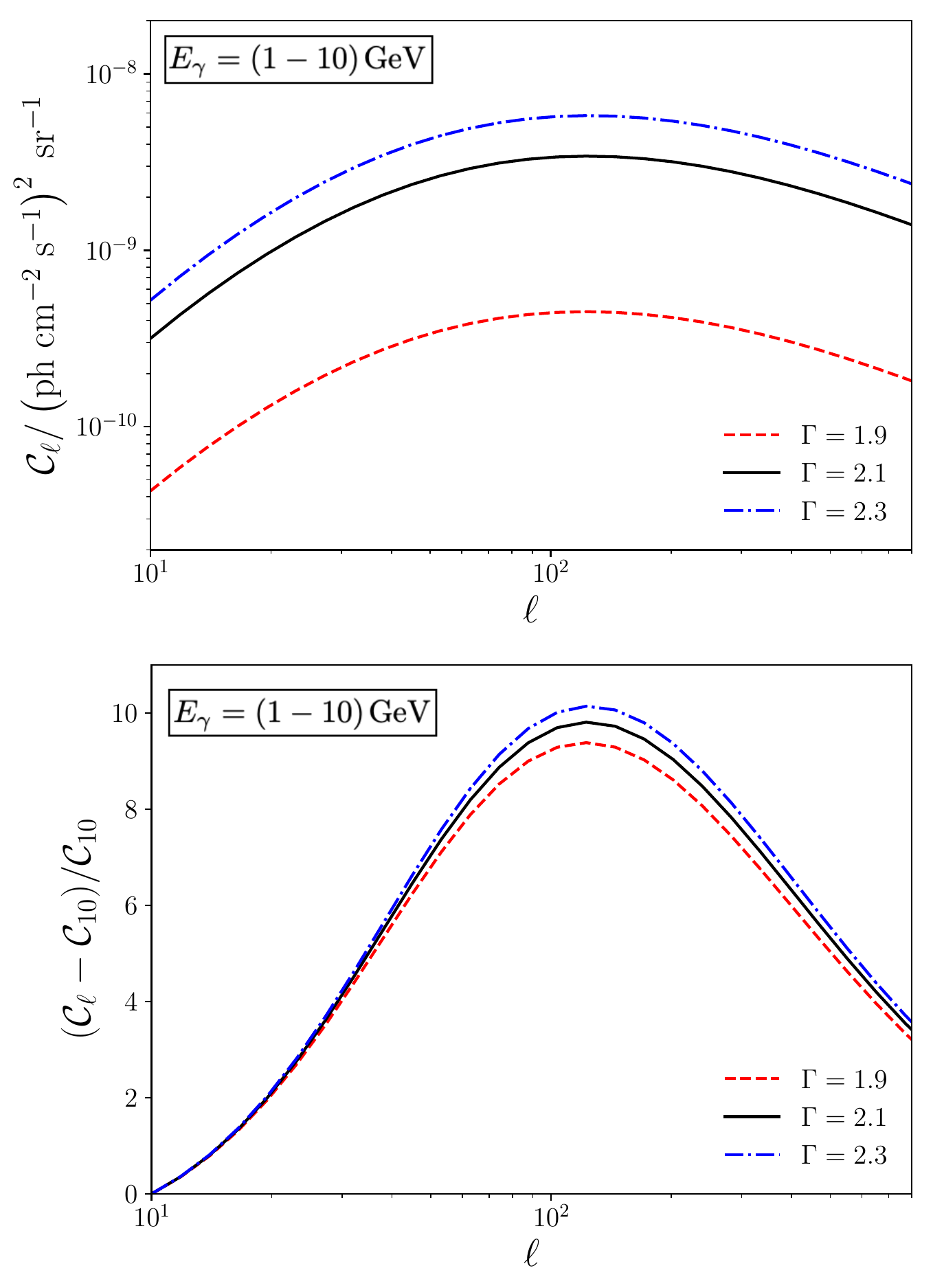}
    \caption{EGB intensity fluctuation angular power spectrum $\mathcal{C}_{\ell}$ shown against multipole $\ell$, for alternative choices of the spectral index of the hadronic CRs, $\Gamma$, in the source population. Top and bottom panels as per Figure~\ref{fig:comparison_energy}. Results are shown for the $\gamma$-ray energy band $E_{\gamma} = (1-10) \;{\rm GeV}$.}
  \label{fig:comparison_powerlaw}
\end{figure}

A similar comparison for variation of $\Gamma$ is shown in Figure~\ref{fig:comparison_powerlaw}, where values of $\Gamma = 1.9$ and 2.3 are considered alongside the fiducial choice. These represent a less-steep (steeper) internal proton spectrum in the SFG population (respectively), as may arise from a younger (older) CR spectrum, or due to variations in accelerator geometries/configurations or CR acceleration physics, and reflects the range of values determined from observational analyses of nearby $\gamma$-ray emitting SFGs~\citep{Ajello2020ApJ_SFG}. The impact of this variation is a change in $\gamma$-ray flux (and hence $\mathcal{C}_{\ell}$ normalisation), as shown in the upper panel, with an increased EGB intensity for a steeper choice of CR index. As the $\gamma$-ray emission spectrum from the SFG closely reflects the hadronic CR spectrum, a steeper CR spectral index yields more power in the $\gamma$-ray energy spectrum at lower energies. From Figure~\ref{fig:spec_and_attenuation}, it can be seen that the strongest attenuation from the source galaxy is felt by higher energy $\gamma$-rays, so the fraction attenuated within SFGs is reduced for steeper CR spectral indices. 
The lower panel of Figure~\ref{fig:comparison_powerlaw} reveals the shape of the EGB anisotropy power spectrum is also influenced by the choice of $\Gamma$, where a steeper CR spectrum yields a noticeably sharper EGB angular power peak, while a softer CR spectrum produces a broader peak. This effect follows from the energy dependence of $\gamma$-ray attenuation in the EBL: despite the internal attenuation, a less steep CR source spectrum would ultimately still produce a higher fraction of high-energy $\gamma$-rays. These are attenuated more readily, and fewer photons from distant sources survive to $z=0$, even when considering the cascade process. The fraction of flux contributed by SFGs at large distances (which would imprint signatures on larger angular scales) is therefore reduced for less steep CR spectra, effectively suppressing EGB angular power, particularly on larger scales. Recent work has considered the possibility of blended spectral indices within SFGs~\citep{Ambrosone2021MNRAS}. The results here would imply these would have a non-trivial impact on the EGB anisotropy, and should be explored further in future studies.
%Variation of parameters pertaining to the ISM of source SFGs (including density, multi-phase structure, chemical composition and dust abundance/distribution) could also impact these results. Exploration of these quantities falls beyond the scope of this work and is left to future studies -- in this first approach, the parametric mean treatment adopted here is sufficient for our purposes.

\begin{figure}
    \centering
    \includegraphics[width=\columnwidth]{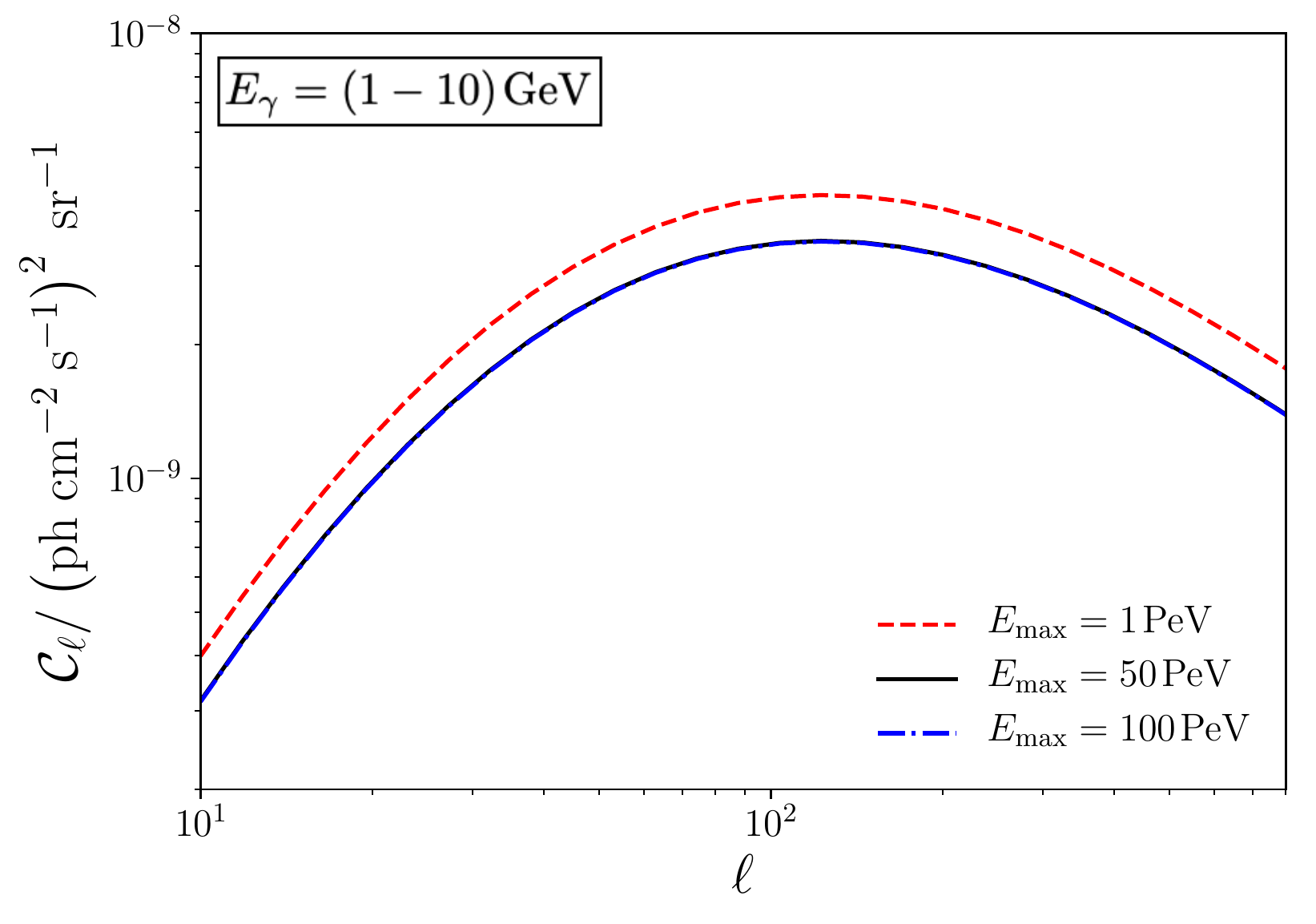}
    \caption{EGB intensity fluctuation angular power spectrum $\mathcal{C}_{\ell}$ shown against multipole $\ell$, for alternative choices of maximum CR energy, $E_{\rm max}$. The impact on the intensity shown here results from the dependence of the CR spectral normalisation on $E_{\rm max}$ (see equation~\ref{eq:cr_norm}), and is not physical. There is no impact on the shape of the EGB anisotropy. Results are shown for the energy band $E_{\gamma} = (1-10) \;{\rm GeV}$.}
  \label{fig:upper_energy_compare}
\end{figure}

The upper limit of the CR spectrum in SFGs is determined by acceleration mechanisms and the detailed configuration of the accelerators~\citep[e.g. for discussion, see][]{Peretti2020MNRAS}, and the exact value that should be adopted in any given environment remains unsettled. 
However, we find this is not of particular consequence to our results. Figure~\ref{fig:upper_energy_compare} considers alternative choices of $E_{\rm max}$, which shows a limited effect on the EGB intensity. Only a small intensity boost is seen if a lower maximum cut-off is adopted, or a proportionally small decrease arises if a higher cut-off is instead chosen. This can be predominantly accounted for by the adjustment in the spectral normalisation for different choices of $E_{\rm max}$  (see equation~\ref{eq:cr_norm}), rather than any physical process. The EGB anisotropy is not dependent on the exact choice of $E_{\rm max}$.

\subsubsection{Alternative redshift evolution scenarios}

\begin{figure}
    \centering
    \includegraphics[width=\columnwidth]{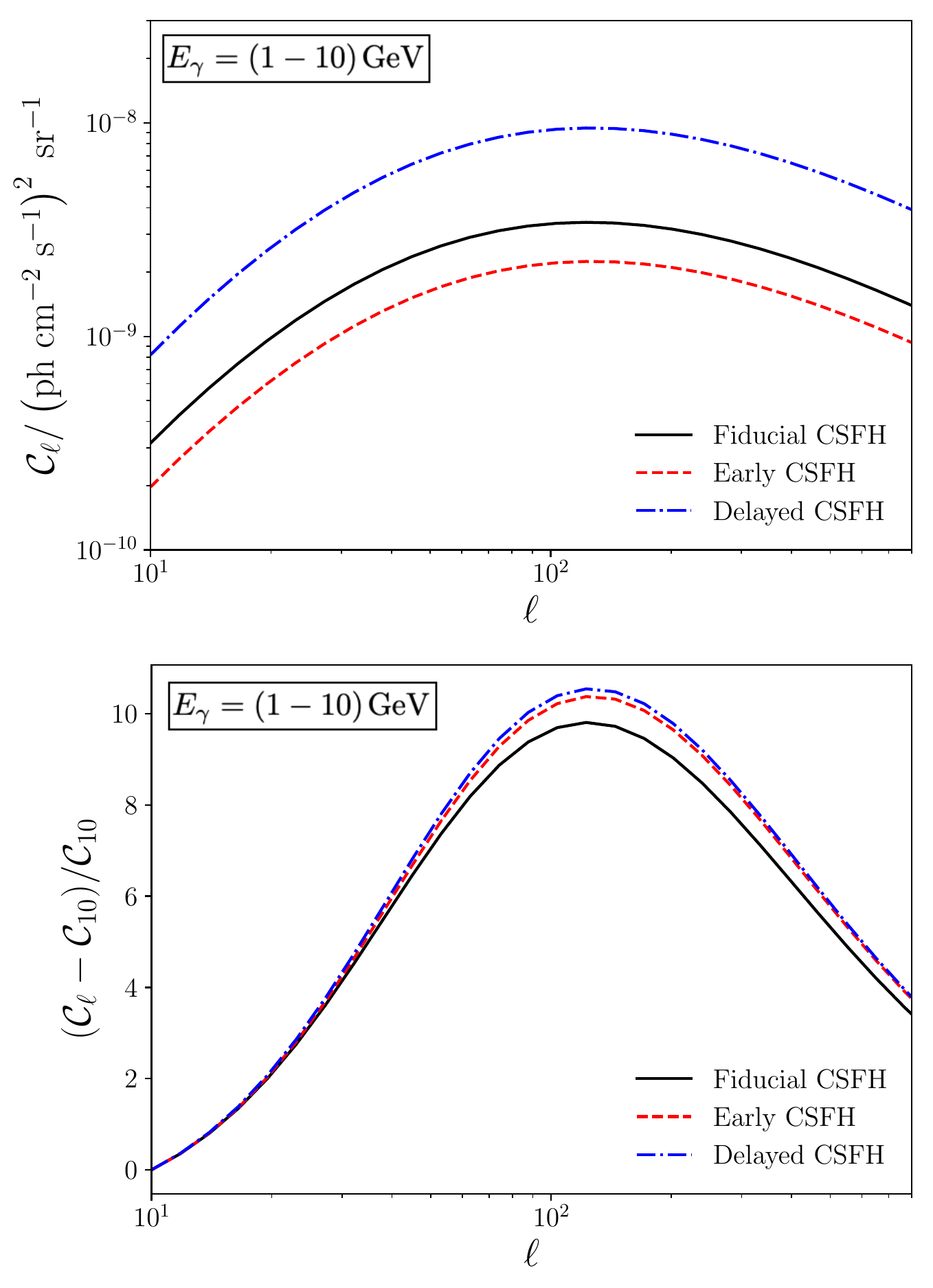}
    \caption{EGB intensity fluctuation angular power spectrum $\mathcal{C}_{\ell}$ shown against multipole $\ell$, for alternative choices of the underlying cosmic star-formation history (CSFH). The fiducial result of Figure~\ref{fig:baseline_anisotropies} is shown by the black solid line, which assumes that the redshift distribution function of SFGs follows~\citealt{Katsianis2017MNRAS}. The redshift distribution adjusted by $\pm 0.5$, yielding `early' and `delayed' CSFH models, as shown. Results are shown for the energy band $E_{\gamma} = (1-10) \;{\rm GeV}$.}
  \label{fig:redshift_evo_compare}
\end{figure}

%May also be modified by gravitiational lensing...? Brings stuff from higher-z to have a greater contribution than it otherwise would...

Our fiducial model adopts the galaxy population model of~\citealt{Katsianis2017MNRAS}, which yields a redshift distribution of cosmic star-formation broadly compatible with~\cite{Madau2014ARAA}, where the peak of cosmic star-formation arises at $z\sim2$. However, this may not fully reflect the diverse redshift distributions of various classes of SFGs (e.g. the distribution of sub-mm galaxy samples in~\citealt{Simpson2014ApJ} compared to that of the luminous sub-mm sources in~\citealt{Koprowski2014MNRAS} or dusty star-forming galaxies in~\citealt{Strandet2016ApJ}), which are not guaranteed to follow the global mean cosmic star-formation history (CSFH) of the Universe.
We crudely demonstrate the level of impact alternative CSFHs would have on the EGB in Figure~\ref{fig:redshift_evo_compare}, where we modify our fiducial distribution derived from~\citealt{Katsianis2017MNRAS} by simply adjusting its redshift distribution by $\pm 0.5$, thus creating an `early' CSFH model, and a `delayed' CSFH model. The main impact of this is on the EGB intensity, which is reduced for the earlier CSFH model, or increased for the later one (see Figure~\ref{fig:redshift_evo_compare}). This follows largely from our crude adjustment, in that more stars would form in the `early' CSFH scenario (and conversely, fewer in the `delayed' CSFH). However, more subtle effects emerge in the EGB angular power spectrum (Figure~\ref{fig:redshift_evo_compare}, lower panel). It is not intuitive that the spectral shape is broadened both in the `early' and `delayed' CFSH scenarios compared to the fiducial model, with a slightly greater skew towards more power at larger $\ell$s (smaller scales). These can both be understood from the interplay between the redshift distribution of sources in a spherical volume, and the attenuation of $\gamma$-rays in EBL radiation fields: in the `early' CSFH model, there are more sources at higher redshift (imprinting EGB signatures on larger angular scales). However, the greater distance to these sources means a greater degree of $\gamma$-ray attenuation in the intervening EBL, so their contribution (per source) to the $z=0$ EGB would be relatively weak. The is partially compensated by the larger number of sources contained within the volume to a higher redshift, thus broadening the EGB anisotropy signature slightly more over a wider range of scales compared to the fiducial model - i.e. making it less strongly peaked. The converse is true for the `delayed' CSFH model, but the effect is broadly the same due to the EBL attenuation and source distribution acting antagonistically. 

While these crude variations in CSFH offer little physical insight into the astrophysics of SFG populations, they do illustrate that signatures imprinted by SFGs are influenced by their redshift distribution, and that both its intensity and anisotropy encode information about this. EGB anisotropies particularly offer potential as a diagnostic tool to distinguish between different redshift distributions of source populations and, hence, offer scope as a probe the evolutionary histories of population classes of SFGs in which CR activity is important. However, we have shown that these signatures can be subtle, and must be carefully modelled and understood before they can be reliably used to probe and interpret CR activity within source distributions over redshift. 

\subsection{Observational prospects}

\subsubsection{Statistical error}

\begin{figure}
    \centering
    \includegraphics[width=\columnwidth]{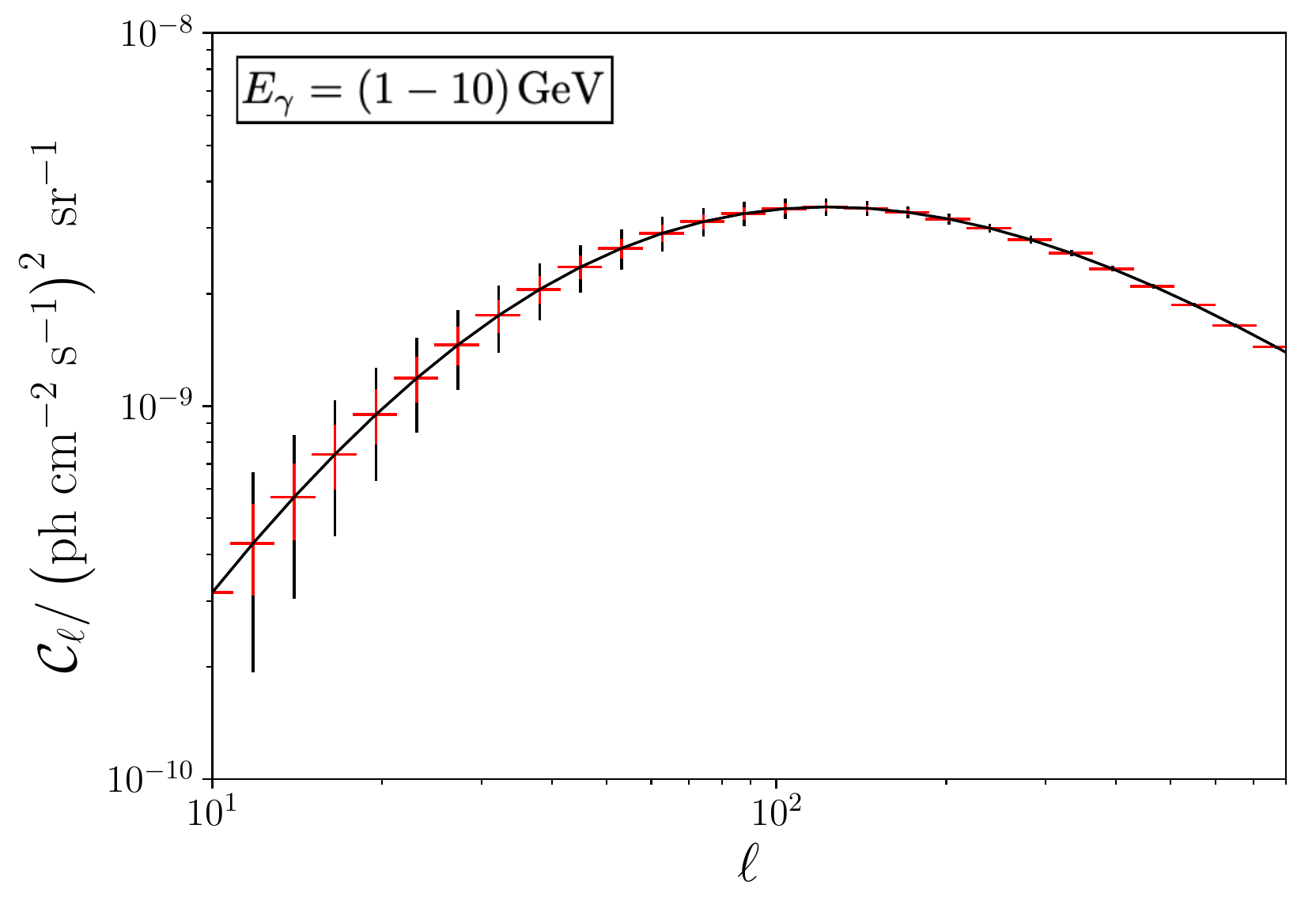}
    \caption{EGB intensity fluctuation angular power spectrum $\mathcal{C}_{\ell}$ for the fiducial model shown against multipole $\ell$ in the energy band $E_{\gamma} = (1-10)\;{\rm GeV}$. Error bars indicate the projected 1-$\sigma$ statistical errors, if considering full sky coverage with $f_{\rm sky} = 1$ in red (reflective of the full-sky coverage of \textit{Fermi}-LAT), or $f_{\rm sky} = 0.25$ in black (indicative of the anticipated sky coverage in CTA's Extra-galactic Survey Key Science Project).}
  \label{fig:baseline_errors}
\end{figure}

The projected statistical 1-$\sigma$ error in an extracted measurement of $\mathcal{C}_{\ell}$ is given by
\begin{equation}
\delta \mathcal{C}_{\ell}^2 = \frac{2 \mathcal{C}_{\ell}^2}{(2\ell + 1) \;\! \Delta \ell \;\! f_{\rm sky}} \ ,
\end{equation}
\citep{Ando2007PhRvD, Ando2007MNRAS}, where $\Delta \ell$ is the bin size in multipole space, and $f_{\rm sky}$ is the fraction of sky covered by the relevant $\gamma$-ray survey. We find this error dominates over all uncertainties built into our model, and would be the primary limitation in resolving EGB signatures.
We show this projected statistical error in $\mathcal{C}_{\ell}$ for 40 equal bins in log $\ell$ space for $f_{\rm sky} = 0.25$ in black (this is indicative of the sky coverage anticipated as part of CTA's Extra-galactic Survey Key Science Project -- see~\citealt{CTA2019_book} for details), and $f_{\rm sky} = 1$ in red (reflective of the full-sky coverage of \textit{Fermi}-LAT) in Figure~\ref{fig:baseline_errors}. It is evident that low multi-poles, or large scale anisotropies are most affected by statistical fluctuations. At intermediate and small scales, the statistical error is greatly reduced (smaller scale anisotropies are computed by splitting the sky into a larger number of regions, thus reducing statistical variations), with good prospects for signal extraction.

%Figure~\ref{fig:comparison_z_errors} demonstrates the level to which alternative redshift scenarios may be resolved, given the projected statistical error in an extracted signal (cf. Figure~\ref{fig:redshift_evo_compare}). In this simple scenario, it can be seen that the SFG population peaking at $z^{\star} = 2.5$ can be resolved, but those following distributions peaking at $z^{\star} = 1.5$ and 2.0 are not distinguishable from one another. The separation between the evolutionary scenarios may be somewhat improved by increasing the multipole bin size, however this would reduce the accuracy to which the $\mathcal{C}_{\ell}$ peak can be resolved. 
%\begin{figure}
%    \centering
%    \includegraphics[width=\columnwidth]{cls_z_resolve.pdf}
%    \caption{Uncertainties propagated to normalised anisotropy curves to demonstrate the degree to which three different redshift evolutionary scenarios could be discerned. \textcolor{green}{These all assume that the redshift distribution function of SFGs follows the fitted form in~\citealt{Madau2014ARAA}, but with the peak location adjusted to $z^{\star} = 1.5, 2.0$ and 2.5, as shown.}}
%  \label{fig:comparison_z_errors}
%\end{figure}

\subsubsection{Integration time}

The integration time required to detect EGB anisotropies can be estimated by comparing the $\gamma$-ray background intensity with instrument sensitivity. 
To demonstrate the prospects for detecting a signal, we consider the sensitivities of the current \textit{Fermi}-LAT observatory\footnote{\textit{Fermi}-LAT top-level Pass 8 performance information is available online, see \url{https://www.slac.stanford.edu/exp/glast/groups/canda/lat_Performance.htm}. Instrument response functions are based on the new event analysis and selection criteria described by~\citealt{Atwood2013arXiv}.} and those estimated for the up-coming CTA,\footnote{CTA instrument response functions are provided by the CTA Consortium and Observatory, see \url{http://www.cta-observatory.org/science/cta-performance/} (version prod3b-v2) for more details.} at 50 GeV, where reasonable comparison may be made between the two instruments, and compute the corresponding integration time for our fiducial model in each case. 
The mean intensity of the EGB at 50 GeV was found to be $9.9\times 10^{-8} ~{\rm GeV}\;\!{\rm cm}^{-2}\;\!{\rm s}^{-1}\;\!{\rm sr}^{-1}$ in our fiducial model (see Figure~\ref{fig:EGB_spectrum}), with anisotropic variations leading to minimum intensities on the scales of interest reaching around 5\% of this value. As such, we argue it would be necessary to detect EGB intensities as low as $5.0\times 10^{-9} ~{\rm GeV}\;\!{\rm cm}^{-2}\;\!{\rm s}^{-1}\;\!{\rm sr}^{-1}$ to be able to clearly recover anisotropy signatures. This corresponds to a flux threshold of $7.9\times 10^{-12} ~{\rm erg}\;\!{\rm cm}^{-2}\;\!{\rm s}^{-1}$, which would be detectable with \textit{Fermi}-LAT (at a 5$\sigma$ level and with at least 10 counts per bin) after around 5.5 years of observation. This estimate assumes uniform sky exposure. In reality, the exposure of \textit{Fermi}-LAT varies by a factor of 0.57~\citep[e.g.][]{Nolan2012ApJS}, and so our estimate should be correspondingly increased to at least 10 years for a signature to be observed. 
The 10-year Pass 8 release of LAT data is therefore already reaching sufficient (or near-sufficient) exposure on many scales to detect anisotropies attributed to SFG populations, and accordingly initial detections of EGB anisotropy signatures from SFG populations are emerging~\citep{Fornasa2016PhRvD, Ackermann2018PhRvL}. These will improve over time, as \textit{Fermi}-LAT integration time continues to increase.
The projected CTA integration time to detect the same intensity would be around 0.5 hours (this is approximated from the estimated sensitivity of the CTA-North array at a 70-degree elevation angle). Given that the proposed CTA extra-galactic Survey Key Science Project would cover around 25\% of the extra-galactic $\gamma$-ray sky with a uniform integration time of 1.11 hours~\citep{CTA2019_book},  EGB intensities around 2 times fainter could be reached, depending on the final array configuration and observational strategy adopted. This greatly improves the potential for resolving $\gamma$-ray signatures, and opens the prospect for much more detailed signature extraction.

\subsubsection{Signal contamination}

The results presented in this paper demonstrate how idealised SFG populations could imprint signatures into an idealised EGB. However, prospects for extracting such signatures from real future EGB data is strongly coupled to our understanding 
of contamination arising from imprints from other source populations, or processes which distort the signal of interest.
%One example is bright point sources that are nearby, yet fall below the detection threshold of current instruments and remain unresolved. These could introduce anisotropies in the $\gamma$-ray background -- particularly on small scales -- increasing uncertainties for large multipoles. Removal of most source candidates is possible with point source catalogues, e.g. that released by the \textit{Fermi}-LAT collaboration, periodically~\citep{Ajello2020ApJ}. However, further mitigation efforts in future pipelines may also make use of known source locations (e.g. AGN) in other wavebands to find and mask any $\gamma$-ray excess in their vicinity.
Chief among these is the AGN contribution to the $\gamma$-ray background. Below 0.1 GeV, AGN may account for a fraction of $50^{+12}_{-11}\%$ of the EGB, with around 70\% of this emission having already been resolved by \textit{Fermi}-LAT~\citep{Ajello2015ApJ} -- but this is thought to rise to $85^{+15}_{-21}\%$ at higher energies~\citep{Ajello2015ApJ}; see also~\citep{Inoue2009ApJ, Abdo2010ApJ, Singal2012ApJ, Ajello2014ApJ}. However, the redshift evolutions of AGN populations are not expected to be the same as for SFGs~\citep[see, e.g.][]{Jacobsen2015MNRAS}. This means that their EGB signatures would be imprinted at different length-scales to SFGs and, if sufficiently understood, could be distinguished from SFGs in a high-resolution $\mathcal{C}_{\ell}$ spectrum.

A further source of contamination would arise from the effects of large-scale magnetic fields, which are thought to permeate the Universe. These would have a deflective effect on pair-produced electrons in the 
$\gamma$-ray cascade. Although weak (constraints from non-detections limit their strengths to below $10^{-16} - 10^{-13}$ G; see~\citealt{Han2017ARAA}), the cumulative deflection of a $\gamma$-ray cascade beam over cosmological distances could be sufficient to form a 
broadened halo~\citep{Aharonian1994ApJ}, with its angular spread determined by the strength and structure of the intervening large-scale magnetic fields.
This would distort the original EGB intensity patterns attributed to astrophysical sources, smearing out signals and reprocessing them to different scales (according to the underlying structure of the magnetic field). Indeed, we estimate that the impact of a uniform intergalactic magnetic field 
of strength as low as $10^{-17.5}~{\rm G}$ 
would modify the signature imprinted by our fiducial model 
of between a factor of $\sim$2 (at small scales) and almost an order of magnitude at large scales $\ell\sim10$, if adopting the beam broadening approach of~\citet{Ichiki2008ApJ}.
The exact distortion pattern that would likely arise in real data would presumably be non-trivial and complicated to properly model, given that intergalactic magnetic fields have varying structure and strengths on many scales, and would also evolve over redshift. 
As such, the nature of these distortions must be properly understood and carefully modelled for meaningful interpretations of EGB anisotropies to be possible.

\section{Summary \& Conclusions}
\label{sec:section5}

\begin{figure*}
    \centering
    \includegraphics[width=0.6\textwidth]{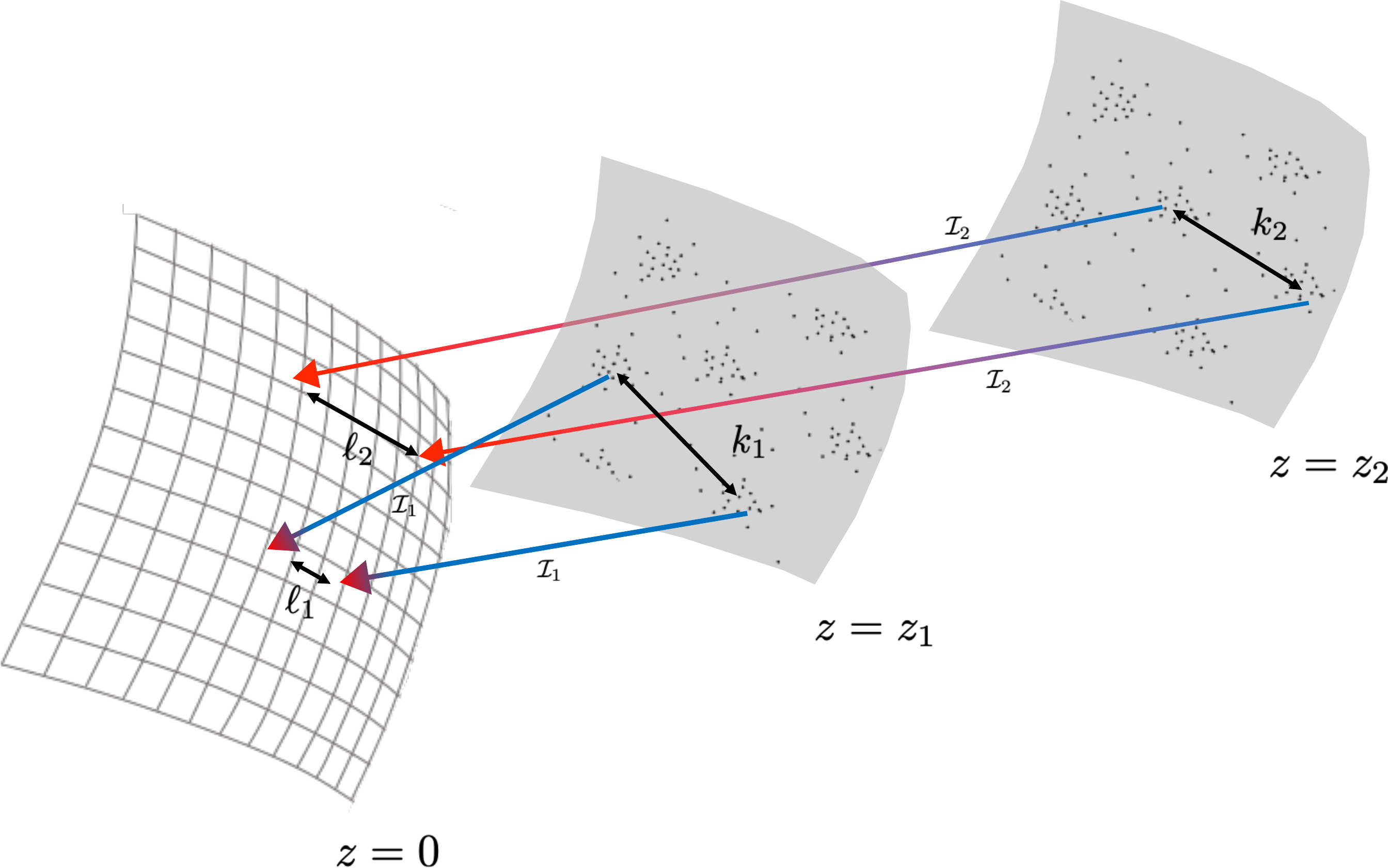}
    \caption{Schematic to illustrate the development of EGB signatures from an evolving unresolved source population. Two redshift slices at $z_1$ and $z_2$, where $z_2>z_1$ are shown, in addition to the $\gamma$-ray sky (grid) at $z=0$. The separation scale $k_2$ at $z_2$ imprints a signature of intensity $\mathcal{I}_2$ with characteristic separation $\ell_2$ on the EGB formed at $z=0$. At $z_1$, the separation scale $k_1$ imprints a signature of intensity $\mathcal{I}_1$ with characteristic separation $\ell_1$, where $\ell_1<\ell_2$. The signatures are reprocessed by the EBL as it propagates through cosmological space, with red colouring indicating an increased proportion of secondary cascaded $\gamma$-rays. The cascade fraction is greater for a beam originating at higher redshift, but the reprocessing does not affect the signal anisotropy pattern directly unless magnetic fields are invoked.}
  \label{fig:EGB_schematic}
\end{figure*}
This work has shown how signatures are imprinted into the EGB by SFG populations (illustrated by the schematic in Figure~\ref{fig:EGB_schematic}), and demonstrated how their contribution may be characterised using 
a small number of physically-motivated parameters. Moreover, it has outlined the relevant EGB statistics that can be used to probe the evolution of the underlying source populations, and has provided a proof-of-concept example by showing the EGB anisotropic signature expected to arise from a population of SFGs.
%\textcolor{green}{following the redshift evolution of~\citet{Madau2014ARAA}.}
This signature is dominated by the contribution from SFGs around the so-called
`high noon' of star-formation at redshifts z$\sim$2-3, where physical conditions and processes in galaxies differ dramatically from those in the local Universe. The interactions of CRs, their associated production of particles and radiation, and their deposition of momentum during this epoch become important factors in controlling the evolution of galaxies and producing energetic cosmic backgrounds. The EGB offers scope to probe these interactions in a direct way, and analysis of patterns within the EGB offer potential 
 to advance our understanding of critical aspects of CR interactions in and around SFGs, in particular during the cosmic noon.

We have further shown that different sub-populations of SFGs could be resolved by a careful analysis of the EGB intensity and angular power spectrum, once appropriate physical models and signal extraction techniques are developed, and that \textit{Fermi}-LAT will soon reach sufficient integration times for signatures imprinted by SFGs to be extracted. This will be substantially improved by up-coming facilities, e.g. CTA, which will offer far greater sensitivities and will be able to resolve SFG source populations in even more detail.
We have also demonstrated that intergalactic magnetic fields can distort imprinted EGB signatures, but the magnitude and structure of this distortion is currently unclear. It is essential that this is explored carefully in future work to ensure that physical interpretations of EGB anisotropies can be reliably made.

In the coming decade, a wealth of new EGB data will become available to the $\gamma$-ray community, with current and up-coming instruments offering unprecedented sensitivities and resolution. There is great potential to use this data to infer new information about cosmic star-formation, intergalactic magnetic fields, SFGs and their properties, and AGN -- if appropriate models for the detailed signatures these imprint in the EGB are available. However, 
theoretical and methodological frameworks must first be urgently developed
to ensure efforts in the community are able to make optimal use of this up-coming data and the opportunities it presents.

\section*{acknowledgements}

This work used high-performance computing facilities operated by the
Center for Informatics and Computation in Astronomy (CICA) at National
Tsing Hua University (NTHU). This equipment was funded by the Ministry of
Education of Taiwan and the Ministry of Science and Technology of Taiwan.
We are also grateful to the National Center for High-performance 
                       Computing (Taiwan) for computer time and facilities.
ERO is supported by the Ministry of Education of Taiwan at CICA, NTHU.
His visits to Kavli IPMU were hosted by KGL, and supported by a travel grant of University College London's Mullard Space Science Laboratory (UCL/MSSL), and the Ministry of Science and Technology of Taiwan. KGL acknowledges support from JSPS KAKENHI Grants JP18H05868 and JP19K14755. 
This research has made use of the CTA instrument response functions provided by the CTA Consortium and Observatory, see \url{http://www.cta-observatory.org/science/cta-performance/} (version prod3b-v2) for details.
ERO thanks Prof. John Silverman (IPMU) and Prof. Yoshiyuki Inoue (Osaka) for helpful discussions about star-forming galaxies and the cosmological propagation and attenuation of $\gamma$-rays, Prof. Masahiro Teshima for facilitating visits to the Institute of Cosmic Ray Research, University of Tokyo, where useful discussions took place, and Prof. John S. Gallagher III (UW-Madison) and Prof. Tomotsugu Goto (NTHU) for comments about the implications of this work. The authors thank 
Prof. Kinwah Wu (UCL/MSSL) 
%, Dr. Alvina Y. L. On (UCL/MSSL), and 
%Y. X. Jane Yap (NTHU) 
for discussions that helped to inform and inspire the early stages of this work, and the anonymous referee for their detailed review which substantially improved the manuscript. This research used NASA's Astrophysics Data Systems.

%\end{acknowledgements}
%%%%%%%%%%%%%%%%%%%%%%%%%%%%%%%%%%%%%%%%%%%%%%%%%%

\section*{Data Availability}

No new data were generated or analysed in support of this research.

%-------------------------------------------------------------------

\bibliographystyle{mnras} % style aa.bst
\bibliography{references} % your references Yourfile.bib

%%%%%%%%%%%%%%%%%%%%%%%%%%%%%%%%%%%%%%%%%%%%%%%%%%

%%%%%%%%%%%%%%%%% APPENDICES %%%%%%%%%%%%%%%%%%%%%

\appendix

\section{Cosmic ray proton density}
\label{sec:cr_proton_density}

\noindent
Strong star-forming activity gives rise to frequent SN events and remnants shortly after the onset of star-formation. Such environments can accelerate particles to relativistic energies 
  through, e.g. \citet{Fermi1949PhRv} acceleration
  in diffusive magnetised shocks~\citep{Axford1977ICRC, Krymskii1977DoSSR, Blandford1978ApJ, Bell1978MNRASI, Bell1978MNRASII} -- see \citet{Blasi2011crpa} for further discussion.
  The energy injected into the relativistic particles (i.e. CRs) is governed
by the total energy provided in the SN events, and the efficiency of acceleration.
To describe their total luminosity, we adopt the relation
\begin{equation} 
   L_{\rm CR, eff} = \varepsilon f_{\nu} E_{\rm SN} {\cal R}_{\rm SN} 
     =   \alpha \left[\frac{\varepsilon f_{\nu} E_{\rm SN} \mathcal{R}_{\rm SF}}{M_{\rm SN}}  \right]   \ .  
     \label{eq:cr_luminosity}
\end{equation}
\citep{Owen2018MNRAS}, where 
     $E_{\rm SN}$ is the SN event total energy (around $10^{53}$ erg for core-collapse Type II P SNe expected to dominate in these highly star-forming systems), 
     $\varepsilon$ is the CR acceleration efficiency (0.1 is set as a conservative value in this work, although some variation could be justified -- see~\citealt{Fields2001A&A, Strong2010ApJ, Lemoine2012A&A, Caprioli2012JCAP, Dermer2013A&A, Morlino2012A&A, Wang2018MNRAS} for indications on the possible range) 
     while $f_{\nu} = 0.01$ is the fraction of energy retained by the SN event after neutrino losses (although we adopt a value of 1\% here, the exact choice would depend on the SN type and environment with fractions as low as 0.1\% being plausible -- for discussion, see  
\citealt{Iwamoto2006AIPC, Smartt2009ARAA, Janka2012ARNPS}). 
     $\alpha$ indicates the fraction of stars which evolve to produce a core-collapse SN event, and this is governed by the initial stellar mass function (IMF) and its upper cut-off. For a Salpeter IMF with index $\Upsilon = 2.35$ between $1\;\!{\rm M}_{\odot}$ and an upper cut-off mass for stars able to ultimately produce a SN event of $M_{\rm SN}  = 50\;\!{\rm M}_{\odot}$~\citep[see, e.g.][]{Fryer1999ApJ, Heger2003ApJ}, this fraction takes a value of $\alpha  = 0.05$, which we fix for all galaxies in the present study~\citep[this was also adopted in earlier work, e.g.][]{Owen2018MNRAS}, assuming no variation in the stellar mass function between galaxies. While future work may consider alternative mass function models and upper values for the mass cut-off, their bearing on the CR flux is relatively weak. $\mathcal{R}_{\rm SN}$ and $\mathcal{R}_{\rm SF}$ are the SN event rate and star-formation rate of a galaxy, respectively. Either may be used to parameterise the CR flux normalisation, with the former being more closely related to the energy-injection rate (and hence of closer bearing to the physical model), while the latter may be considered a closer relation to observational quantities. 
     Hereafter, we specify the CR luminosity of galaxies in terms of $\mathcal{R}_{\rm SF}$ and leave all other parameters in equation~\ref{eq:cr_luminosity} fixed at the stated values.
     
The CR proton density at some location ${\bf r}$ within a galaxy
may be modelled as the superposition of the contributions from an ensemble of 
$N$ continuously-injecting discrete CR sources located at the points described by the position vectors ${\bf r}_{\rm i}$,
\begin{equation}
\label{eq:proton_solution}
n_{\rm p}(\gamma_{\rm p}, {\bf r})\;\!{\rm d}\gamma_{\rm p} = \sum_{i=1}^{N} \; \;\!\frac{f_{\rm adv}\;\! Q_{\rm p}(\gamma_{\rm p}, {\textbf r}_i) \;\! \mathcal{V}_{\rm S} \; \mathcal{A}(\gamma_{\rm p}, {\textbf r}; {\textbf r}_i)}{4\pi r' D(\gamma_{\rm p}, r')} {\rm d}\gamma_{\rm p}\ ,
\end{equation}
\citep{Owen2019AA},
once a galaxy has settled into a steady-state with a roughly constant star-formation rate and saturated magnetic field.  This is typically reached a few tens of Myr after the onset of star-formation for SFGs, with the magnetic saturation time being inversely proportional to $\mathcal{R}_{\rm SF}$~\citep{Schober2013A&A}. $D$ is the energy-dependent CR diffusion coefficient (see section~\ref{sec:diff_param} for details), which takes the parametric form
\begin{equation}
D(\gamma_{\rm p}) = D_0 \left[ \frac{r_L(\gamma_{\rm p},\langle |B| \rangle)|}{r_{L,0}}\right]^{\varsigma} \ ,
\label{eq:parametric_diffusion_coeff}
\end{equation}
where $\langle |B| \rangle = |B|$ is the characteristic interstellar magnetic field strength. The normalisation value $D_0 = 3.0\times 10^{28}$ cm$^2$ s$^{-1}$ is based on empirical measurements of the diffusion of CRs in the interstellar medium (ISM) of the Milky Way and is appropriate for a 1 GeV CR proton diffusing through a 5$\mu$G interstellar magnetic field with corresponding Larmour radius $r_{L, 0}$.
 In equation~\ref{eq:proton_solution}, $f_{\rm adv} = 0.5$ is the fraction of CRs that would be removed from the ISM of a SFG by advection in galactic outflows. Large-scale galactic outflows would be common in distant, young SFGs (see, e.g.~\citealt{Frye2002ApJ, Ajiki2002ApJ}), being driven by the confluence of feedback from the concentrated starburst episode arising in galactic cores~\citep[see][for details about how these may be driven]{Yu2020MNRAS}. For plausible outflow velocities, CR advection timescales would indicate that a substantial fraction of CRs could be removed from the nucleus of a SFG by an outflow wind. Due to the prevalence and strength of outflows in the core of SFGs, we adopt a fiducial value of 50\% here~\citep[as indicated by the timescales shown in][]{Peretti2019MNRAS}, however we consider the exact fraction is unsettled and would vary substantially between different galaxies and model configurations -- e.g. \citet{Owen2019MNRAS} estimated a value of $f_{\rm adv}\sim 0.1$, but calculated this as a fraction of CRs lost from the entire host galaxy, not just the starburst core.
 $\mathcal{V}_{\rm S}$ is the volume of each of the $N$ sources (if chosen physically, this would correspond to the a characteristic size of a SN remnant), and
 $\mathcal{A}$ is an CR attenuation factor due to their interaction losses within the interstellar medium of the host galaxy. This is written as
\begin{equation}
\mathcal{A}(\gamma_{\rm p}, {\textbf r}; {\textbf r}_i) = \exp \left\{ - \int_{{\textbf r}_i}^{{\textbf r}} \varrho(\gamma_{\rm p}, {\textbf r}') \; \alpha^*(\gamma_{\rm p}, {\textbf r}') \; {\rm d}{\textbf r}' \right\} \ ,
\label{eq:proton_atten}
\end{equation}
 which quantifies the level of attenuation experienced by a beam of CR protons between a source at location ${\textbf r}_{i}$, and some general location ${\textbf r}$ at a distance of $r' = |{\textbf r}_{i}-{\textbf r}|$ apart. Here, $\alpha^*(\gamma_{\rm p}, {\textbf r}') =  n_{\rm H}({\textbf r}')\;\!\sigma_{\rm p\pi}(\gamma_{\rm p})$, and
 $\varrho$ is
the ratio of the free-streaming (i.e. the attenuation length due to CR interactions $\ell_{\rm p\pi} = 1/\alpha^*$ in a non-magnetised medium) and diffusive path lengths of the CRs,
\begin{equation}
\varrho(\gamma_{\rm p},  {\textbf r}') = \frac{\ell_{\rm p\pi}}{\ell_{\rm diff}} = \left\{\frac{{c}}{4 \; D(\gamma_{\rm p}) \; n_{\rm H}({\textbf r}') \; {\sigma}_{\rm p\pi}(\gamma_{\rm p})}\right\}^{1/2} \ ,
\end{equation}
\citep{Owen2018MNRAS}.
The term $Q_{\rm p}$ in equation~\ref{eq:proton_solution} quantifies the CR injection rate discretised by source, such that $Q_{\rm p}\;\!{\rm d}\gamma_{\rm p}$ is the rate of injection of CRs within an energy interval ${\rm d}\gamma_{\rm p}$. We define this as
\begin{equation}
Q_{\rm p}(\gamma_{\rm p}, {\textbf r}_i) = S_{\rm N}(r)\;\!\frac{\mathcal{L}_{0}}{N}\;\! \frac{\partial}{\partial \gamma_{\rm p}} \left(\frac{\gamma_{\rm p}}{\gamma_{\rm p, 0}}\right)^{-\Gamma} \biggr\vert_{\;{\textbf r}_i} \ ,
\end{equation}
  where the volumetric CR injection rate is $S_{\rm N}(r)$, and the normalisation
  \begin{equation}
\label{eq:cr_norm}
   \mathcal{L}_0 = \frac{L_{\rm CR, eff}(1-\Gamma)E_0^{-\Gamma}}{E_{\rm max}^{1-\Gamma} - E_0^{1-\Gamma}}  \ , 
\end{equation}
follows from the CR energy budget set by equation~\ref{eq:cr_luminosity}.
Here, we use a reference energy $E_0 = \gamma_{\rm p, 0} m_{\rm p} c^2$ of 1 GeV, and set a maximum CR energy of $E_{\rm max} = 50$ PeV~\citep{Peretti2019MNRAS}. We relax this choice in section~\ref{sec:model_params}, where we consider the impact of different choices of maximum CR energy.

This aim of this study is to assess the contribution of galaxies to the $\gamma$-ray background.
As such, a detailed formulation of the sub-galactic variations in CR density
is not required. Instead, we set $S_{\rm N} = 1$ and so approximate the CR proton density $n_{\rm p}$  to be uniform through each galaxy interior (correspondingly, we also consider a uniform gas density, of mean value $n_{\rm H} = \langle n_{\rm H}\rangle = 1~{\rm cm}^{-3}$, and leave more detailed considerations of the impact of the ISM density and structure to future work). 
This also removes the need for discretisation of the model, so we set $N=1$ in equation~\ref{eq:proton_solution}, which reduces to
\begin{align}
n_{\rm p}(\gamma_{\rm p})\;\!{\rm d}\gamma_{\rm p} &= \frac{R^3 \;\! f_{\rm adv}\;\!\mathcal{L}_{0} \; \mathcal{A}(\zeta_{\rm p\pi})}{3 \langle{r'}\rangle D(\gamma_{\rm p})} \;\! \frac{\partial}{\partial \gamma_{\rm p}} \left(\frac{\gamma_{\rm p}}{\gamma_{\rm p, 0}}\right)^{-\Gamma} \;\!{\rm d}\gamma_{\rm p} \ , \nonumber \\
& = \; \frac{35 R^2 \;\! f_{\rm adv}\;\!\mathcal{L}_{0} \;\! \mathcal{A}(\zeta_{\rm p\pi})}{108 \;\! D(\gamma_{\rm p})} \;\! \frac{\partial}{\partial \gamma_{\rm p}} \left(\frac{\gamma_{\rm p}}{\gamma_{\rm p,0}}\right)^{-\Gamma} \;\!{\rm d}\gamma_{\rm p} \ .
\label{eq:final_protons_eq}
\end{align}
Here $\mathcal{A}(\zeta_{\rm p\pi})$ is the mean attenuation of protons as they propagate through the host galaxy (the form of the mean attenuation function $\mathcal{A}(...)$ for a uniform sphere is given by equation~\ref{eq:final_mean_atten}), and $\langle{r'}\rangle = 36 R/ 35$ is the mean separation length within a spherical volume to radial points uniformly distributed throughout the volume, where $R$ sets the characteristic size of the volume -- in this case, the radius of a characteristic star-forming region of the host galaxy.
%(this would be comparable to the size of the starburst nuclei considered in~\citealt{Peretti2020MNRAS}). 
We also define 
$\zeta_{\rm p\pi} = (\varrho\;\!R/\ell_{\rm p\pi})^{1/2}$, and $\ell_{\rm p\pi} = ({\sigma}_{\rm p\pi} n_{\rm H})^{-1}$ is the mean free path of protons undergoing hadronic interactions as they propagate through a uniform ISM of density $n_{\rm H}$.

\subsection{CR diffusion parameter in star-forming galaxies}
\label{sec:diff_param}

Equation~\ref{eq:parametric_diffusion_coeff} sets $D$ as the diffusion coefficient for CRs in the ISM of their host galaxy. It is specified by both the CR energy, $\gamma_{\rm p}$ and the ambient mean magnetic field strength, $\langle |B| \rangle = |B|$. For such an environment, we set the normalisation value as $D_0 = 3.0\times 10^{28}$ cm$^2$ s$^{-1}$, which is based on empirical measurements of the diffusion of CRs in the ISM of the Milky Way. It would be appropriate for a 1 GeV CR proton diffusing through a 5$\mu$G interstellar magnetic field with corresponding Larmour radius $r_{L, 0}$. We argue that there is no strong physical basis to motivate different values in star-forming galaxies, where the processes which set this value are not likely to be different to those seen in the local Universe, and consider that alternative values would not imply more correct physics. $\varsigma$ is introduced in equation~\ref{eq:parametric_diffusion_coeff} to encode the interstellar magnetic turbulence. For this, we adopt a value of 1/2~\citep[e.g.][]{Berezinskii1990book, Strong2007ARNPS}, which is appropriate for a Kraichnan-type turbulence spectrum following a power law of the form $P_t(k) \; {\rm d} k \approx k^{-2+\varsigma}$, and is thought to be a reasonable description for the turbulence in an ISM~\citep[see][]{Yan2004ApJ, Strong2007ARNPS}. 

The magnetic field of the host galaxy is also thought to be driven by star-formation via a turbulent dynamo mechanism during a starburst phase~\citep[see e.g.][]{Beck2012MNRAS, Latif2013MNRAS, Schober2013A&A}, and observational studies favour the rapid development of magnetic fields in protogalaxies,
 reaching strengths comparable to the Milky Way within a few Myr of their formation~\citep{Bernet2008Nat, Beck2012MNRAS, Hammond2012arXiv, Rieder2016MNRAS, Sur2018MNRAS}. 
  The saturation level of such a mechanism, for example that introduced by~\citet{Schober2013A&A}, may be approximated by invoking equipartition with the kinetic energy of the turbulent gas, $B_{\rm L,sat} =\left[{4 \pi \; \rho}\right]^{1/2} v_{\rm f} f_t$ where 
  $\rho = m_{\rm p}\;\!n_{\rm H}$ is the local gas density and $v_{\rm f}$ is the fluctuation velocity ($v_{\rm f}  \approx R_{\rm gal} ({2}\pi \rho G/3)^{1/2}$ for the protogalaxy, if adopting a pressure with gravity equilibrium approximation -- see~\citealt{Schober2013A&A}). Here, $f_t$ represents the deviation from exact equipartition to account for the efficiency of energy transfer from the turbulent kinetic energy to magnetic energy, which simulation work estimates to be around 10\%~\citep[see, e.g.][]{Federrath2011PRL, Schober2013A&A}.
  
\section{Attenuation of $\gamma$-rays in matter and radiation fields}
\label{sec:atten_comp}

Section~\ref{sec:internal_attenuation} considers that $\gamma$-ray absorption 
in SFGs is predominantly attributed to $\gamma\gamma$ pair production in ambient radiation fields. 
However, other studies have argued that 
dense interstellar clouds may also have a role.
In \citet{Lacki2012arXiv}, the ability of $\gamma$-rays to ionise
dense interstellar clouds was discussed. 
Beams of hadronic CRs can easily undergo interactions in dense environments, where substantial attenuation may arise, leading to the production of leptons, neutrinos and $\gamma$-rays. A similar mechanism invoking CR beam dumping is presented in~\citet{Vereecken2020arXiv}.
The pp inelastic cross section (equation~\ref{eq:pp_cs}) is 
of order $\sigma_{\rm p\pi} \sim 10^{-26}\;\!{\rm cm}^2$, being only weakly dependent on CR energy. The corresponding CR path length due to pp losses in the dense core of a molecular cloud with volume density of $n_{\rm H} = 10^5 \;\! {\rm cm}^{-3}$ would be $\ell_{\rm p\pi} \approx \left(\sim n_{\rm H} \sigma_{\rm p \pi} \right)^{-1} \sim 300 \;\!{\rm pc}$. This is substantially larger than the size of a dense core (which would typically extend for less than a pc -- see, e.g.~\citealt{Bergin2007}), suggesting CR attenuation in such an environment would be relatively unimportant. 
Nevertheless, $\gamma$-ray production would still arise, with a fraction of around 0.1 \% of the CR beam intensity undergoing hadronic interactions through the dense core of a molecular cloud. Around 1/3rd of the attenuated CR energy would then be passed to $\gamma$-rays, as can be determined from the branching ratios of the intermediate pions~\citep{Dermer2009_book}. 
These $\gamma$-rays could be attenuated by Bethe-Heitler $\gamma Z$ pair production ($\gamma + Z \rightarrow e^- + e^+ + Z$, where $Z$ is an atomic nucleus -- see~\citealt{Lacki2012arXiv, Vereecken2020arXiv}).
In inner regions of SFGs, dense clouds are common: high volume filling fractions of clouds in star-forming nuclei have been inferred for nearby luminous infrared galaxies such as Arp 299~\citep{Sliwa2012ApJ}, and inner gas volume densities in some regions of such systems could exceed $10^{5}\;\!{\rm cm}^{-3}$~\citep[e.g.][]{Imanishi2019ApJS}. Taking an extreme interpretation of these findings to estimate an upper limit for the attenuative effect of 
dense clouds on $\gamma$-rays in a SFG nucleus,
we consider a uniform density medium of $10^{5}\;\!{\rm cm}^{-3}$ throughout a $R = 0.1~{\rm kpc}$ nuclear starburst region.
At 1 GeV, the Bethe-Heitler $\gamma Z$ pair production cross section is $\sigma_{\rm A\gamma} \sim 10^{-26}\;\!{\rm cm}^2$~\citep{Berestetskii1980_book}, and is not strongly energy-dependent.  This gives a $\gamma$-ray attenuation fraction of around 30\%, with a mean path-length of $\ell_{\rm A\gamma} \sim 0.3~{\rm kpc}$. 
Comparing with Figure~\ref{fig:spec_and_attenuation} (upper panel), this process would dominate over losses due $\gamma\gamma$ interactions up to around 10 GeV.
However, in the lower panel, it can be seen that a 30\% reduction in CR flux below 10 GeV would not be of great consequence to the emitted spectrum. As such, we do not consider $\gamma Z$ losses in SFG nuclei in this work, and leave this to more detailed future studies.

\section{Internal attenuation of $\gamma$-rays in SFGs}
\label{sec:appendix_a}

The attenuation of $\gamma$-rays within a SFG nucleus can be characterised as an average mean attenuation through a homogeneous, isotropic spherical volume.
In this scenario,
the distance of the path between two points with positions defined by the vectors ${\bf p}_1$ and ${\bf p}_2$ is given by
\begin{equation}
s^2({\bf p}_1, {\bf p}_2) = r_1^2 + r_2^2 -2r_1 r_2 \cos \psi
\end{equation}
where $ \cos \psi = \mu_1 \mu_2 + (1-\mu_1^2)^{1/2}(1-\mu_2^2)^{1/2} \cos (\phi_1-\phi_2)$ is their angular separation, and $\mu_1$ and $\mu_2$ are the cosines of their angular positions, $\theta_1$ and $\theta_2$ respectively. The attenuation of a $\gamma$-ray beam along each of the separation paths then follows as $A({\bf p}_1, {\bf p}_2) = \exp[-s({\bf p}_1, {\bf p}_2)/\ell]$ which, for an ensemble of points ${\bf p}_i$ following a random uniform distribution, may be written as
\begin{align}
\langle A({\bf q}) \rangle & = \frac{\int_{V_i} \;\! A({\bf q}, {\bf p}_i) \;\! {\rm d}{V_i}}{\int_{V_i} {\rm d} {V_i}} \nonumber \\
& = \frac{3}{4\pi R^3} \;\! \oint_{\Omega_i} {\rm d}\Omega_i \;\! \int_{r_i=0}^R \exp\left(-s({\bf q}, {\bf p}_i)/\ell\right) \;\! r_i^2 \;\! {\rm d}r_i \;\! \ ,
\end{align}
where ${\rm d}V_i = r_i^2\;\!{\rm d}r_i\;\!{\rm d}\Omega_i$ is the differential element corresponding to the volume $V_i$ occupied by the points distribution ${\bf p}_i$ in the continuous limit.
Under spherical symmetry, the mean attenuation between some reference position ${\bf q}$ at a radius $r$ within the sphere to the points in the ensemble ${\bf p}_i$ (again, in the continuous limit) follows by averaging over $S_r$, the spherical surface specified by $r$, i.e:
\begin{align}
\langle A(r) \rangle & = \frac{\int_{S_r} \langle A({\bf q}) \rangle \;\! {{\rm d}S_r}}{\int_{S_r} {{\rm d}S_r}} \nonumber \\
& = \frac{2}{8\pi R} \oint_{\Omega} {{\rm d}\Omega} \; \frac{\partial}{\partial r} \int_{r'=0}^{r} \langle s({\bf q}) \rangle \;\!r'\;\! {\rm d}r'
\label{eq:general_mean_surf}
\end{align}
where the surface element is separated into its radial and angular components, for which ${\rm d}S_r = 2 r' {\rm d}r' \;\! {\rm d}\Omega$. This may be discretised and reduced into a numerical Monte-Carlo problem by distributing a uniform random ensemble of $N = 10^6$ points within a sphere of radius $R$ to calculate the mean attenuation $\langle A(r) \rangle$ between the points and a radial line $r$ from the origin $r=0$ to the edge of the sphere $r=R$. 
The total characteristic attenuation through the sphere then follows by taking the mean value of $A(r)$ along the radial line from $r = 0$ to $R$, i.e:
\begin{align}
\mathcal{A}(R, \ell) & = \frac{1}{R}\;\!\int_{r=0}^R A(r) \;\! {\rm d}r \ .
\label{eq:main_mc}
\end{align}
The resulting characteristic attenuation is then specified by only the size of the sphere (describing the extent of the nuclear core of a star-forming galaxy) and the effective mean path length of the $\gamma$-rays within the sphere's internal medium (density and radiation field), $l_{\rm mfp}$. Equation~\ref{eq:main_mc} may be well-approximated by a Gaussian function,
\begin{equation}
\mathcal{A}(\zeta) = \exp\left(-{\zeta^2}\right) \ ,
\label{eq:final_mean_atten}
\end{equation}
where $\zeta = (R/l_{\rm mfp})^{1/2}$. We find this approximation gives an error of less than 1\% for all reasonable values of $\zeta$.
%fit (eq.~\ref{eq:final_mean_atten}) and residual in Figure~\ref{fig:comparison_mc}, where it can be seen that the resulting error is negligible (at worst, less than 1\% for all reasonable values of $\zeta$).
%\begin{figure}
%    \centering
%    \includegraphics[width=\columnwidth]{attenuation_approx_mcfit.pdf}
%    \caption{{\bf Top panel}: mean attenuation computed by the Monte-Carlo method outlined in the text (points), and corresponding fitted function~\ref{eq:final_mean_atten}. {\bf Bottom panel}: residuals show the fit is better than 1\% for the range of $\zeta$.}
%  \label{fig:comparison_mc}
%\end{figure}

\section{EGB clustering and Poisson noise}
\label{sec:clustering_power_spec}

\noindent
The SFG power spectrum $P_{\rm g}(k, z)$ would imprint a signature in the EGB, and this could be measured from $\gamma$-ray background observations using the auto-correlation function~\citep[e.g.][]{Peebles1980, Inoue2013ApJ_b},
\begin{align}
\label{eq:autocorr_term_app}
    \mathcal{C}(\theta) & = \langle \delta I({\bf l_1}) \;\! \delta I({\bf l_2}) \rangle \\ \nonumber
    & = \frac{1}{16 \pi^2} \int_0^{\infty} {\rm d}l_1\;\!\int_0^{\infty} {\rm d}l_2 \;\! \xi({\bf l_1}-{\bf l_2})\;\! I({\bf l_1}) \;\! I({\bf l_2}) \ ,
\end{align}
where $\xi(...)$ is the two-point correlation function, and $I({\bf l})$ is the intensity of the EGB at some position specified by the co-moving vector ${\bf l}$, of (co-moving) length $l$. $\delta I({\bf l})$ is the intensity fluctuation, being the deviation of $I$ at some position ${\bf l}$ from its mean value, and $\theta$ is the angular separation of the positions ${\bf l_1}$ and ${\bf l_2}$.
In general, the clustering (or correlation) term of the angular power spectrum can be defined as the Fourier Transform of the auto-correlation function. 

In the case of EGB analyses, we consider a $\gamma$-ray signal in a 2-dimensional space $\mathcal{C}_{\ell}^{\gamma}$,  
which can be split into two components,
\begin{equation}
\mathcal{C}_{\ell}^{\gamma} = \mathcal{C}_{\ell}^{P} + \mathcal{C}_{\ell}^{C}
\label{eq:sum_contributions}
\end{equation}
where $\mathcal{C}_{\ell}^{P}$ is an isotropic Poisson noise term (an auto-correlation term), and $ \mathcal{C}_{\ell}^{C}$ is the clustering term between points of angular separation $\theta>0$. Previous work assessing the AGN contribution to the EGB~\citep[e.g.][]{Inoue2013ApJ_b} found the 
Poisson term to be comparable to the clustering term. In this case, the EGB contribution was comprised of a population of bright, unresolved point sources. The SFG contribution would be different, with the $\gamma$-ray background emission being dominated by a much larger number of galaxies, each being fainter than a typical AGN source. As such, the $\mathcal{C}_{\ell}^{P}$ term would presumably be much smaller than the $\mathcal{C}_{\ell}^{C}$ term here.

We compute the Poisson and correlation terms separately.
For the clustering term, we only require (isotropic) angular separations such that it reduces to
\begin{equation}
    \mathcal{C}_{\ell}^{C} = \int_{\theta> 0} {\rm d}^2\theta \;\! e^{-i{\bf l} \cdot {\bf \Btheta}} \;\! \mathcal{C}(\theta) \ ,
\end{equation}
where the separation angle $\theta$ is non-zero. 
Assuming that the mean signal intensity is the same between the positions at ${\bf l_1}$ and ${\bf l_2}$ and that the signal $I$ is statistically isotropic and homogeneous both in space and its projection onto the sphere, and if adopting the notation $r_2 = l_2 - l_1$ and $r_1 = (l_2+l_1)/2$ for convenience (cf. the~\citealt{Limber1953ApJ} approximation), we arrive at
\begin{align}
    \mathcal{C}_{\ell}^{C} & = \frac{1}{16 \pi^2} \;\! \int_{\theta> 0} {\rm d}^2\theta \;\! e^{-i{\bf l} \cdot {\bf \Btheta}} 
    \;\! \int_0^{\infty} {\rm d}r_1
    \;\! \int_{-2r_1}^{2r_1} {\rm d}r_2
    \;\! \xi(r_2 \hat{\bf r} + r_1 \theta \hat{\bf \Btheta}) \;\! I^2 \nonumber \\
    & = \frac{1}{16 \pi^2} \;\! \int_{\theta> 0} {\rm d}^2\theta e^{-i{\bf l} \cdot {\bf \Btheta}} \;\! \int_0^{z_{\rm max}} \frac{{\rm d}^2V_{\rm c}}{{\rm d}z\;\!{\rm d}\Omega}{\rm d}z\;\ \nonumber \\
    & \hspace{2cm} \times \int_{-\infty}^{\infty} {\rm d}r_2 \frac{\xi(r_2 \hat{\bf r} + r_1 \theta \hat{\bf \Btheta}) \;\! L_{\gamma}^2(z)}{r_p^2 (1+z)^2} \ ,
    \label{eq:cls_original}
\end{align}
where we approximate the limits of $r_2$ to be $\pm \infty$. Here, 
$r_p$ is the proper distance in the $\hat{\bf r}$ direction, $V_{\rm c}$ is a comoving volume and $\hat{\bf r}$ and $\hat{\bf \Btheta}$ are introduced as unit vectors in the radial direction towards the background (thus being a function of redshift) and the direction between the two points at ${\bf l_1}$ and ${\bf l_2}$, respectively. $L_{\gamma}(z)$ is the $\gamma$-ray luminosity of the source population at a distance of redshift $z$.
%\footnote{In transforming from the first line to the second line of equation~\ref{eq:cls_original}, we have used the substitution
%\begin{equation}
%{\rm d}r_1 = {\rm d}z \frac{{\rm d}^2 V_{\rm c}}{{\rm d}z {\rm d}\Omega}\;\! \frac{1}{r_p^2 (1+z)^2} \ ,
%\end{equation}
%where $r_1$ is comoving, and $r_p$ is a proper length. We approximate the limits of $r_2$ to be $\pm \infty$.}

The two-point correlation function is related to the underlying power spectrum of the $\gamma$-ray source population by a Fourier Transform,
\begin{equation}
    \xi(r_2 \hat{\bf r} + r_1 \theta \hat{\bf \Btheta}) = \frac{1}{{8\pi^3}}\int {\rm d}^3k \;\! P_{\rm g}(k) \exp\left(i {\bf k}\cdot [r_2 \hat{\bf r} + r_1 \theta \hat{\bf \Btheta}]\right) \ .
\end{equation}
%which we argue would trace the matter power spectrum such that $P(k) = P_{\rm m}(k)$. 
Here, $k$ is a proper wavenumber, defined as the magnitude of the wavevector ${\bf k}$.
We may combine and decompose the exponential terms of equation~\ref{eq:cls_original} into 
\begin{equation}
\exp\left(i {\bf k}\cdot [r_2 \hat{\bf r} + r_1 \theta \hat{\bf \Btheta}]-i{\bf l} \cdot {\bf \Btheta}\right) = \exp\left(i r_2 \;\! {\bf k}_{||}\cdot \hat{\bf r} + i \Btheta \cdot [r_1 {\bf k}_{\perp} - {\bf l}] \right)
\end{equation}
if expressing ${\bf k}$ in terms of parallel and perpendicular components to the direction $\hat{\bf r}$.
The integrals over $r_2$ and $\theta$ in equation~\ref{eq:cls_original} may then be readily evaluated as:
\begin{align}
    \int_{\theta> 0}  \int_{r_2 = -\infty}^{\infty} {\rm d}^2\theta \;\! {\rm d}r_2 \exp & \left( i r_2 \;\! {\bf k}_{||}\cdot \hat{\bf r} + i \Btheta \cdot [r_1 {\bf k}_{\perp} - {\bf l}] \right) \nonumber \\
    = & ~\frac{8\pi^3}{r_{\rm p}^2} \delta(k_{||, {\rm p}})\;\!\delta^2\left(r\;\!k_{\perp, {\rm p}} - l_{\rm p}[1+z]\right) \ ,
\end{align}
where the second step converts comoving coordinates to proper coordinates such that $r_{\rm p, 2} = (1+z) \;\! r_2$, and subscript $p$ denotes proper coordinates (as used in the wavenumbers), and the equivalence of the $k_{\perp}$ components follows from the earlier assumption that the signal is isotropic in the sky plane.
Equation~\ref{eq:cls_original} then reduces to
\begin{align}
    \mathcal{C}_{\ell}^{C} = & \frac{1}{16 \pi^2}
    \int_0^{z_{\rm max}} \frac{{\rm d}^2V_{\rm c}}{{\rm d}z\;\!{\rm d}\Omega}{\rm d}z\;\ \frac{L_{\gamma}^2(z)}{r_p^2\;\!(1+z)^2} \nonumber \\
    & \hspace{0.5cm} \times \;\! \int {\rm d}k_{||}\;\!{\rm d}^2k_{\perp}\;\!P(k_{\perp})\;\!\delta(k_{||})\;\! \delta^2(r_p \;\!k_{\perp} - l_p[1+z])\;\!r_p^{-2} \nonumber \\
    & = \frac{1}{16 \pi^2}
    \int_0^{z_{\rm max}} \frac{{\rm d}^2V_{\rm c}}{{\rm d}z\;\!{\rm d}\Omega}{\rm d}z\;\ \frac{L_{\gamma}^2(z)\;\!P(l_p[1+z]/r_p)}{r_p^4\;\!(1+z)^2} \ ,
\end{align}
after integrating over the orthogonal wave-vector components, 
which is the combined contribution of SFGs to the EGB up to some redshift $z_{\rm max}$. In differential units of flux, this gives
\begin{equation}
\mathcal{C}_{\ell}^{C}(E_{\gamma}) =  \int_0^{z_{\rm max}} \frac{{\rm d}^2V_{\rm c}}{{\rm d}z\;\!{\rm d}\Omega}{\rm d}z\;\ P\left(\frac{\ell_p}{r_p}[1+z]\right)\left\{ \frac{{\rm d}F_{\gamma}(E_{\gamma}, z)}{{\rm d}E_{\gamma}} \right\}^2
%\frac{{\rm d}\mathcal{C}_{\ell}^{\gamma}(E_{\gamma})}{{\rm d}E_{\gamma}} =  \int_0^{z_{\rm max}} \frac{{\rm d}^2V_{\rm c}}{{\rm d}z\;\!{\rm d}\Omega}{\rm d}z\;\ P\left(\frac{\ell}{r_p}\right)\left\{ \frac{{\rm d}F_{\gamma}(E_{\gamma}, z)}{{\rm d}E_{\gamma}} \right\}^2
\label{eq:differential_anisotropy_app}
\end{equation}
which is equation~\ref{eq:differential_anisotropy} in the main text.
%\footnote{In equation~\ref{eq:f_z_func} the energy $E_{\gamma}$ of the photon has been red-shifted by a factor $1+z$, and the leading factor of $1+z$ is from bandwidth compression, with photons emitted over the energy range $(1+z)\Delta E_{\gamma}$ being squeezed into the energy range $\Delta E_{\gamma}$ in the observer's frame.} 
Here,
\begin{equation}
\frac{{\rm d}F_{\gamma}(E_{\gamma}, z)}{{\rm d}E_{\gamma}} = \frac{1+z}{4\pi D_{\rm L}^2}\frac{{\rm d}L_{\gamma}^{\rm Tot}(E_{\gamma}[1+z], z)}{{\rm d}E_{\gamma}} \ ,
%\frac{{\rm d}F_{\gamma}(E_{\gamma}, z)}{{\rm d}E_{\gamma}} = \frac{1+z}{4\pi D_{\rm L}^2}\frac{{\rm d}L_{\gamma}^{\rm Tot}(E_{\gamma}[1+z], z)}{{\rm d}E_{\gamma}}
\label{eq:f_z_func}
\end{equation}
which describes the redshift-dependent emission of $\gamma$-rays from the source population of SFGs, thus incorporating the internal and external $\gamma$-ray attenuation/reprocessing models, and the co-moving number density of SFGs.
$D_{\rm L}$ is the luminosity distance, 
defined as
\begin{equation}
D_{\rm L} = (1+z) \left(\frac{c}{H_0}\right)\int_0^z \frac{{\rm d}z'}{E(z')} \ .
\label{eq:lum_dist_app}
\end{equation}
for $E(z) = \left[\Omega_{\rm m,0}(1+z)^3 + \Omega_{\rm \Lambda,0} + \Omega_{\rm r,0}(1+z)^4\right]^{1/2}$, 
where terms retain their earlier definitions.
%with $\Omega_{\rm m,0} = 0.315\pm0.007$, $\Omega_{\rm r,0} \approx 0$ and $\Omega_{\rm \Lambda,0} = 0.685\pm0.007$ as the normalised density parameters for matter, radiation and dark energy respectively, and $H_0 = 100 h ~\text{km}\;\!\text{s}^{-1}\;\!\text{Mpc}^{-1}$ as the present value of the Hubble constant, where $h = 0.673\pm 0.006$~\citep{Planck2018}.
The Poisson term $\mathcal{C}_{\ell}^{P}$ is then given by
\begin{equation}
\mathcal{C}_{\ell}^{P}(E_{\gamma}) =  \int_0^{z_{\rm max}} \frac{{\rm d}^2V_{\rm c}}{{\rm d}z\;\!{\rm d}\Omega}{\rm d}z\;\! \left\{ \frac{{\rm d}F_{\gamma}(E_{\gamma}, z)}{{\rm d}E_{\gamma}} \right\}^2 \ , 
\label{eq:final_poisson_app}
\end{equation}
which is equation~\ref{eq:final_poisson} in the main text.

\section{Computational method}
\label{sec:computational_method}

\noindent
We directly compute the EGB anisotropy statistic at $z=0$ for an energy $E_{\gamma}$ using equation~\ref{eq:differential_anisotropy}. 
To do this, we adopt a numerical approach where the volume to redshift $z_{\rm max}$ containing the EGB source population is discretised into $N_{\rm z}$ shells. The contribution from each shell to the EGB at $z=0$ is calculated by solving equation~\ref{eq:radiative_transfer} subject to the boundary condition set by the $\gamma$-ray intensity at the originating shell. This is the combined contribution of the (primary) SFG $\gamma$-ray emission from the shell, taken through the volume between shell $i$ and $i+1$, plus a (secondary) background contribution to that shell -- i.e. the propagated emission from higher redshift shells and their cascaded component reprocessed to an energy $E_{\gamma}$ (equation~\ref{eq:emission_function}).

The primary $\gamma$-ray emission of the SFG galaxy population at each shell is found by integrating the $\gamma$-ray contribution of a galaxy per star-formation rate (computed from the SFRF -- see section~\ref{sec:section3} for details) between 1 and 10000 $\text{M}_{\odot}\;\!\text{yr}^{-1}$, using a logarithmic trapezium-rule with $N_{\rm SFR}$ steps. The original $\gamma$-ray spectral emissivity per galaxy (equation~\ref{eq:emiss_spec}) is computed in the same manner, using $N_{\rm g}$ steps. The primary $\gamma$-ray contribution is convolved with the SFG power spectrum to encode the spatial dependence of the emission. 

The secondary $\gamma$-ray emission function in the transfer equation~\ref{eq:emission_function} depends on the electron injection rate, given by equation~\ref{eq:electrons_rate}. This, in turn, is set by the absorption of $\gamma$-rays propagating to a shell $i$ from more distant shells. The integral in equation~\ref{eq:electrons_rate} must therefore be evaluated for each step, $i$. This is achieved by a second level of discretisation, with a further redshift grid defined from $j = 1$ to $N_{\rm z, 2}$, with the primary $\gamma$-ray intensity computed at each `sub'-shell (again, using equation~\ref{eq:radiative_transfer}). At each sub-shell, the contribution from the full spectra of higher-redshift $\gamma$-rays must be considered, as a fraction of the spectrum at each energy will provide a contribution to lower energy $\gamma$-rays, via the cascade process. The double integral of~\ref{eq:emission_function} is computed by discretisation of the dimensionless variable $x_c$ into $N_{x_c}$ steps (for the inner integral), and the outer integral follows likewise with $N_{\rm \gamma}$ steps. We found a simple trapezium-rule numerical integration method to be sufficient for both of these. 

The inner redshift integral could be computed by a simple adaptive Runge-Kutta (RK) Fehlberg 4th order scheme~\citep{Press1992_book}, while we required the greater numerical stability afforded by an implicit RK 4/5 scheme for the outer redshift grid. For this, we used the \verb RADAU5 \; solver of~\cite{Hairer1993book}. 
%We note that the algorithm described above scales as $\mathcal{O}(N_{z}^4)$. A nested parallelization scheme over the outer redshift grid (main level), and the inner grid over $x_c$, cf. equation~\ref{eq:emission_function} (sub level), was possible and improved this to $\mathcal{O}(N_{z}^3)$, with a maximum speed-up factor of up to $N_{x_c}$.
Strictly, a redshift grid with a finer resolution than the typical absorption length of $\gamma$-rays in the EBL should be adopted. However, we found that sub-grid variations in the $\gamma$-ray intensity due to this attenuation and cascade re-emission made little difference to our results. As such, a coarser grid could be safely adopted, and a choice of $N_{\rm z} = 100$ and $N_{\rm z, 2} = 100$ was found to give results which varied by less than 1\% compared to higher resolution grids (we compared to grid resolutions increased by a factor of two in both cases). To achieve a comparable 1\% level of numerical accuracy, we found minimum resolutions of $N_{x_c} = 10$ and $N_{\rm \gamma} = 10$ were sufficient to numerically evaluate the integrals in equation~\ref{eq:emission_function}. Higher resolution discretisation was found to be necessary for the source function $\gamma$-ray emission computations (sections~\ref{sec:gamma_ray_from_gal} and~\ref{sec:sfrf}, respectively), with both $N_{\rm SFR} = 50$ and
$N_{\rm g} = 50$.

\bsp	% typesetting comment
\label{lastpage}
\end{document}